\documentclass{article}

\PassOptionsToPackage{numbers, compress}{natbib}

\usepackage[final]{neurips_2025}
\usepackage{wrapfig}
\usepackage{amsmath, amsthm, amssymb, mathtools, listings}

\usepackage[utf8]{inputenc} 
\usepackage[T1]{fontenc}    
\usepackage{hyperref}       
\hypersetup{citecolor=MidnightBlue}
\hypersetup{linkcolor=black}
\hypersetup{urlcolor=MidnightBlue}

\usepackage{url}            
\usepackage{booktabs}       
\usepackage{amsfonts}       
\usepackage{nicefrac}       
\usepackage{microtype}      
\usepackage{xcolor}         
\usepackage{graphicx}

\definecolor{todoColor}{RGB}{255, 0, 0}   

\newcommand{\animalA}{\mathsf{A}}
\newcommand{\animalB}{\mathsf{B}}

\DeclareMathOperator{\corr}{\mathsf{Corr}}
\DeclareMathOperator{\rsa}{\mathsf{RSA}}

\DeclareMathOperator{\rdm}{\mathsf{RDM}}
\DeclareMathOperator{\train}{\mathsf{train}}
\DeclareMathOperator{\test}{\mathsf{test}}
\DeclareMathOperator{\trueA}{t^{\animalA}_{\train}}
\DeclareMathOperator{\trueB}{t^{\animalB}_{\train}}
\DeclareMathOperator{\trueBtest}{t^{\animalB}_{\test}}

\DeclareMathOperator{\trueAid}{t^{\animalA}}
\DeclareMathOperator{\trueBid}{t^{\animalB}}
\DeclareMathOperator{\sfAtrain}{s^{\animalA}_{1,\train}}
\DeclareMathOperator{\ssAtrain}{s^{\animalA}_{2,\train}}
\DeclareMathOperator{\sfBtrain}{s^{\animalB}_{1,\train}}
\DeclareMathOperator{\ssBtrain}{s^{\animalB}_{2,\train}}

\DeclareMathOperator{\sfBtest}{s^{\animalB}_{1,\test}}
\DeclareMathOperator{\ssBtest}{s^{\animalB}_{2,\test}}
\DeclareMathOperator{\sfAid}{s^{\animalA}_{1}}
\DeclareMathOperator{\ssAid}{s^{\animalA}_{2}}

\DeclareMathOperator{\sfBid}{s^{\animalB}_{1}}
\DeclareMathOperator{\ssBid}{s^{\animalB}_{2}}
\newcommand{\Mtrue}{\mathcal{M}_{\text{true}}}
\newcommand{\Mest}{\widehat{\mathcal{M}}_{\text{est}}}

\title{Intrinsic Goals for Autonomous Agents: Model-Based Exploration in Virtual Zebrafish Predicts Ethological Behavior and Whole-Brain Dynamics}

\author{%
Reece Keller$^{1, 2}$\thanks{Corresponding author.} \quad Alyn Kirsch $^2$\quad Felix Pei$^1$  \quad Xaq Pitkow$^{1, 3}$ \\ \quad \textbf{Leo Kozachkov}$^{4,}$\thanks{These authors jointly supervised this work.} \quad \textbf{Aran Nayebi}$^{3,1,2,}$\footnotemark[2] \\
$^1$Neuroscience Institute, Carnegie Mellon University \\ $^2$Robotics Institute, Carnegie Mellon University\\ $^3$Machine Learning Department, Carnegie Mellon University \\ $^4$IBM Thomas J. Watson Research Center, IBM Research\\
\texttt{\{rdkeller, fcpei, akirscht, xpitkow, anayebi\}@andrew.cmu.edu}\\
\texttt{leokoz8@gmail.com}
}

\begin{document}

\maketitle



\begin{abstract}
Autonomy is a hallmark of animal intelligence, enabling adaptive and intelligent behavior in complex environments without relying on external reward or task structure.
Existing reinforcement learning approaches to exploration in reward-free environments, including a class of methods known as \textit{model-based intrinsic motivation}, exhibit inconsistent exploration patterns and do not converge to an exploratory policy, thus failing to capture robust autonomous behaviors observed in animals.
Moreover, systems neuroscience has largely overlooked the neural basis of autonomy, focusing instead on experimental paradigms where animals are motivated by external reward rather than engaging in ethological, naturalistic and task-independent behavior. 
To bridge these gaps, we introduce a novel model-based intrinsic drive explicitly designed after the principles of autonomous exploration in animals. 
Our method (3M-Progress) achieves animal-like exploration by tracking divergence between an online world model and a fixed prior learned from an ecological niche.
To the best of our knowledge, we introduce the first autonomous embodied agent that predicts brain data entirely from self-supervised optimization of an intrinsic goal—without any behavioral or neural training data—demonstrating that 3M-Progress agents capture the explainable variance in behavioral patterns and whole-brain neural-glial dynamics recorded from autonomously behaving larval zebrafish, thereby providing the first goal-driven, population-level model of neural-glial computation.
Our findings establish a computational framework connecting model-based intrinsic motivation to naturalistic behavior, providing a foundation for building artificial agents with animal-like autonomy.

\end{abstract}

\section{Introduction}
\label{sec:intro}
Animals exhibit remarkable autonomy, navigating complex environments through self-directed, internally driven behaviors rather than solely responding to external rewards or immediate physiological needs. Unlike typical artificial agents designed to optimize explicit, predefined task objectives in well-defined problem settings, animals intrinsically explore and adapt in open-ended, naturalistic environments where goals are neither clear nor stable. 
This capacity for autonomous behavior allows organisms, including humans, to flexibly engage in abstract thinking, exploratory learning, and innovative problem-solving. 
Central to this autonomy is the ability to generate intrinsic goals, seek novel experiences, and continuously refine internal models of the world that inform future actions. 
Understanding the interplay between extrinsic and intrinsic motivations remains a fundamental challenge in both systems neuroscience and artificial intelligence, particularly for building robust agents capable of lifelong autonomy.

\begin{figure}
    \centering
    \includegraphics[width=1\linewidth]{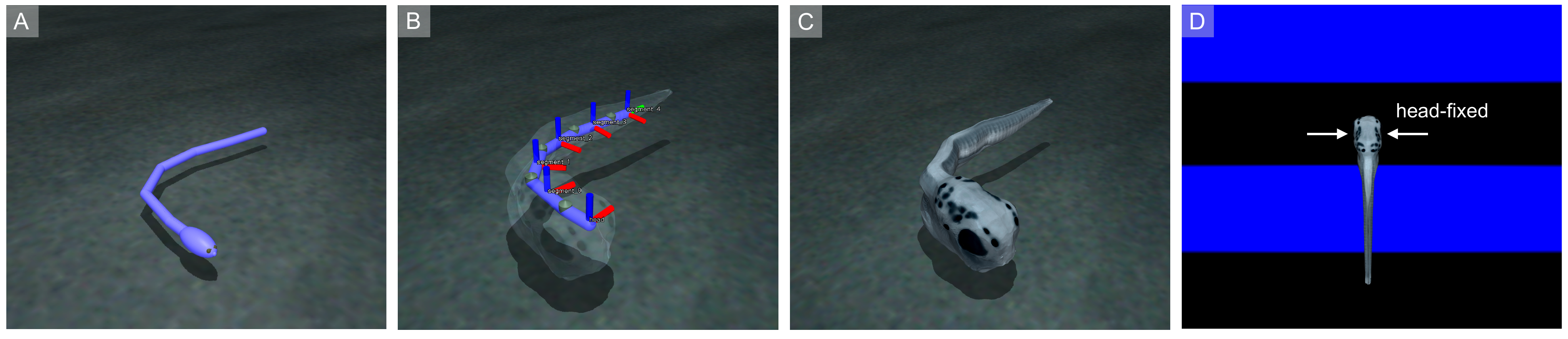}
    \caption{Simulation of the zebrafish agent in a physics-based virtual environment. A) The 6-link embodiment geometry \citep{tassa2018deepmind} in a environment with dynamic fluid forces. B) The agent controls the torque exerted by motors at each joint (5 DoF) to swim and navigate its environment. C) A custom cosmetic skin to mimic the appearance of larval zebrafish. D) A virtual environment matching the experimental parameters of the open loop protocol \citet{mu2019glia}. The root joint located at the head is fixed during training.}
    \label{fig:body-schematic}
    \vspace{-0.1cm}
\end{figure}

However, existing methods in reinforcement learning for intrinsic motivation using world-models
---such as curiosity-driven exploration and variants~\citep{pathak2017curiosity,burda2018exploration,pathak2019self,kim2020active}---exhibit inconsistent behavioral patterns due to difficulties distinguishing between controllable and uncontrollable stimuli, coping with non-stationary environments, and avoiding the ``noisy TV'' problem~\citep{schmidhuber2010formal}. 
For example, agents driven by reinforcing prediction-error alone gets locked into pursuing inherently unpredictable or irrelevant aspects of the environment, hindering their ability to develop robust exploratory behaviors~\citep{pathak2017curiosity}. 
Moreover, existing methods typically
fail to produce structured and stable behavioral transitions that characterize genuine autonomous behaviors observed in animals and children~\citep{haber2018learning}.

At the same time, systems neuroscience has historically overlooked the cellular basis of autonomy, favoring experimental paradigms centered on explicit external rewards to motivate task-dependent behaviors rather than free, unconstrained exploration. As a result, the cellular mechanisms underlying naturalistic autonomy—particularly those involving whole-brain interactions—remain poorly understood. Growing experimental, computational, and theoretical evidence suggests that non-neuronal cells, especially astrocytes, play a critical role in the generation of intelligent behavior \citep{mu2019glia,nagai2021behaviorally, robin2018astroglial, williamson2024learning,kozachkov2023building,kozachkov_pnas_2025,murphy2023conceptual}. 
Owing to their close association with neurons, distinctive connectivity (with a single astrocyte capable of interfacing with up to a million nearby synapses) and ability to perform integrative computations using a hierarchy of timescales, astrocytes are well-positioned to support adaptive, naturalistic, goal-directed behavior \citep{ma2016neuromodulators, stobart2018cortical, doron2022hippocampal, de2023specialized, noh2023cortical, rupprecht2024centripetal}.

\paragraph{A Dataset for Animal Autonomy}To investigate these questions, we study autonomy in larval zebrafish---a uniquely valuable animal model due to their optical transparency, which affords whole-brain calcium imaging via light-sheet microscopy \citep{ahrens2013whole}. Given that the cellular basis of autonomous behavior remains largely unknown, whole-brain imaging allows us to search across the entire recorded population of over 250,000 cells (roughly 125K neurons and 125K astrocytes) to measure how our proposed model aligns with brain activity \citep{mu2019glia} \footnote{The dataset we use is open-source and freely available \href{https://doi.org/10.48324/dandi.000350/0.240822.1759}{here}.}. 
We examine a particular cognitive behavior known as futility-induced passivity, an experimentally induced state of "giving up" observed across the animal spectrum \citep{mu2019glia, brady1958avoidance, camp2015repeated, popovic2023rats}. 
In zebrafish, futility-induced passivity occurs when their swimming effort in a virtual environment fails to simulate optic flow, a key visual feedback signal associated with motor-action under ethological circumstances. After some time in the passive state, zebrafish reattempt to swim--an exploratory behavior to assess whether the open-loop dynamics of the environment have changed. This cycle between active and passive states continues over the course of a trial. 

\citet{mu2019glia} offers rigorous experimental evidence suggesting this transition to passivity is driven neuron-glial interactions in the lateral medulla oblongata (L-MO) that detect and accumulate sensorimotor feedback error. 
Noradrenergic (NE) neurons signal swim failures which slowly activate radial astrocyte with extending processes to the NE cluster of L-MO. At a critical level of intracellular calcium, astrocyte processes downstream of L-MO activate GABAergic neurons that suppress premotor neurons relevant for swimming. 
With extensive manipulation experiments, \citet{mu2019glia} demonstrate that futility-induced passivity is not due to fatigue, struggle, or positional homeostasis. Specifically,  when visual feedback from closed-loop trials was replayed during the open-loop condition, zebrafish still transitioned to a passive state. This indicates the behavior is mediated by an internal model that is used to detect active sensorimotor feedback error, providing strong evidence for model-based intrinsic motivation. 

\paragraph{Contributions} In this work, we study neural-glial computations and their relationship to autonomous animal behavior by training an embodied zebrafish agent with model-based intrinsic motivation and studying its emergent behavior and internal representations. We introduce a novel intrinsic reinforcement learning algorithm termed Model-Memory-Mismatch Progress (3M-Progress) alongside a virtual environment that captures the fundamental physics of the zebrafish embodiment.
Our approach leverages an internal model that continually compares the agent's online memory formed by its current sensory experience against an ethologically relevant prior memory; intrinsic reward then reinforces transitions between behaviors that maximize the divergence between memories relative to its temporal history.
3M-Progress is uniquely capable of producing stable, ethologically-relevant behavioral transitions among several state-of-the-art exploration algorithms in reinforcement learning. 

By training embodied agents with 3M-Progress, we successfully replicate both the robust behavioral patterns and whole-brain neural-glial dynamics in autonomously-behaving zebrafish. 
Capturing nearly all of the variance in neural and astrocytic activity, this marks the first predictive and normative model of neural-glial computation. Our agent was \textit{not trained on any behavioral or neural data}, and thus represents the first autonomous embodied agent that predicts brain data completely from optimizing a self-supervised, intrinsic goal.
To summarize, our technical contributions are: 
\begin{itemize}
    \item 3M-Progress, a novel intrinsic reward that leverages an ecological dynamics prior to guide exploration in new environments. 
    \item Emergence of an autonomous behavior known as futility-induced passivity in 3M-Progress agents, closely matching larval zebrafish behavior. 
    \item Alignment between whole-brain calcium response and 3M-Progress agents, providing the first goal-driven model of neural-glial computation. 
    \item A general modeling perspective positioning intrinsic reinforcement learning as a computational framework for understanding autonomy in animals. 
\end{itemize}
\section{Related Work}
\label{sec:related-work}
\paragraph{Neural-Glial Models}
Although glial cells---especially astrocytes---are increasingly recognized as crucial to adaptive brain function \citep{mu2019glia,nagai2021behaviorally,doron2022hippocampal,murphy2023conceptual,noh2023cortical,rupprecht2024centripetal,williamson2024learning,cagla_science,lefton_science}, computational models of neural-glial interactions remain underdeveloped \citep{kastanenka2020roadmap}. Existing models of neuron-astrocyte dynamics typically fall into two categories: phenomenological models that reproduce specific experimental findings like calcium oscillations or epileptic activity \citep{kozachkov2017causal,de2009glutamate,goldberg2010nonlinear,volman2012computational}, and simplified, ``bottom-up'' mathematical models that explore theoretical principles based on astrocytes’ unique morphology and anatomy \cite{kozachkov2020sequence,ivanov2021increasing,de2022multiple,gong2023astrocytes,kozachkov2023building,kozachkov_pnas_2025}. 
While important, these models are not directly applicable to our setting because they are (a) not yet directly trained on ethological tasks, embodied, or quantitatively validated against real brain data, and (b) typically focused on a single astrocyte rather than a population of astrocytes. 
In contrast, we adopt a ``top-down'' approach: we train a general-purpose recurrent architecture to control an embodied agent to perform ethologically-relevant behavior and find that this imposes strong constraints on the learned representations, allowing us to identify units whose activity patterns closely match those observed in neurons and astrocytes of larval zebrafish. 

\paragraph{Curiosity-driven Exploration} Exploration using self-supervised world-models has demonstrated promising success in several standard reinforcement learning domains, and even more recently in language modeling ~\citep{sun2025curiosity}. Methods like learning progress ~\citep{kim2020active} and Random Network Distillation (RND) ~\citep{burda2018exploration} were primarily evaluated using either handcrafted object-centric or state observations with low-dimensionality embodiments, limiting the applicability to pixel-based environments or continuous control. The Intrinsic Curiosity Module (ICM) ~\citep{pathak2017curiosity} uses a pixel-encoder trained with an inverse-dynamics loss to predict features rather than raw states, but was evaluated on discrete pixel-based environments like Doom and Atari. While some recent works, such as LEXA~\citep{mendonca2021discovering} or Plan2Explore~\citep{sekar2020planning}, extend intrinsic curiosity (specifically, Disagreement ~\citep{pathak2019self}) to continuous control from visual inputs, success of the exploration policy is defined relative to downstream task generalization. Since these methods do not independently evaluate the quality of the exploration policy, it remains unclear whether these algorithms are powerful enough to learn complex behaviors.  
In contrast, we investigate completely open-ended autonomous behavior in reward-free, continuous MDPs with high-dimensional observations, turning the focus on the ability of the exploration algorithm to develop ethological, interpretable behaviors independent of any downstream task. 


\paragraph{Embodied AI in Neuroscience}
Several works have leveraged embodied AI to bridge computational models with neuroscience, including virtual animal models such as the virtual rodent~\citep{Merel2020,aldarondo2024virtual}, which facilitates grounded studies of motor control by replicating rodent motor behaviors across various tasks using imitation learning; the  virtual fruit fly, a biomechanically detailed model matching both the visual system and basic flight capacity used to study a diverse range of behaviors driven by imitation learning~\citep{vaxenburg2024whole, lobato2022neuromechfly, wang2024neuromechfly}; the OpenWorm project~\citep{sarma2018openworm}, a biophysically accurate simulation of the \textit{C-elegan} nematode, but has yet to be combined with deep learning and task-optimization; and Zador et al.'s Embodied Turing Test position paper~\citep{zador2023catalyzing}, which emphasizes developing AI models whose sensorimotor capabilities rival those of their biological counterparts. There are several existing works that apply task-optimization to control details musculoskeletal models \citep{perez2024modeling, codol2024motornet, vargas2024task}, use robots to implement and validate neural circuits in zebrafish \citep{liu2025artificial}, and model animal-like social behavior or object perception using digital twins \citep{wood2024digital, mcgraw2024parallel}.

Our work extends these directions by providing the first predictive and normative computational formalization of neural-glial interactions in embodied agents, thereby validating a circuit model recently proposed by~\citet{mu2019glia}. Most importantly, our model is trained entirely via intrinsically-motivated exploration, unlike previous approaches that constrain behavior and their resulting neural representations by supervised learning ~\citep{Merel2020, aldarondo2024virtual, vaxenburg2024whole}. 
To the best of our knowledge, this marks the first completely autonomous, embodied agent model of behavioral and brain data in neuroscience, pointing towards a promising computational framework for understanding naturalistic, task-independent behavior in biological systems using intrinsic reinforcement learning. 

\section{Methods}
\label{sec:methods}
\paragraph{Virtual Zebrafish Environment}
Animals are physically coupled to the environment through their embodiment; this coupling is often referred to as the sensory-motor or perception-action feedback loop. Physical embodiment imposes strong constraints on both the sensory stream from which an agent learns meaningful representations of the world and the actuation system by which the agent manifests behavior. 
Following this top-down view of biological systems, we construct an embodied agent and custom virtual environment in the MuJoCo physics engine~\citep{todorov2012mujoco} specifically designed after the ethology of the zebrafish (Figure \ref{fig:body-schematic}). 
Leveraging the procedurally generated $n$-link swimmer and built-in inertial fluid model from the Deepmind Control Suite (dm-control) ~\citep{tassa2018deepmind}, we construct an ethological environment (Figure \ref{fig:body-schematic}A-C) in which the agent can freely behave in the presence of both passive and active fluid currents, similar to the dynamic water environments to which zebrafish are native. 
To evaluate our agent in the futility-induced passivity task, we construct a second environment closely matching the open-loop experimental protocol in \citet{mu2019glia} (Figure \ref{fig:body-schematic}D). 
In this environment, agents passively experience a high-contrast grating moving away from the egocentric point-of-view, which simulates backward motion. 
In the closed-loop condition, the agent is head-free and can learn a positional-homeostasis policy to counteract the perceived backward flow. 
In the open-loop condition, the agent is head-fixed and its swim commands produce no movement. 
The passive speed of the moving grating, its colors, and sizing relative to the zebrafish body were determined from experimental parameters in \citet{mu2019glia}. 
These environments provide sensory and actuator configurations that closely match the basic structure of free swimming as well as the head-fixed protocol, facilitating a meaningful comparison between artificial and biological agents.

\paragraph{Agent Design} 
The zebrafish agent is equipped with a recurrent sensory-cognitive architecture to support perception and action in continuous, high-dimensional environments. Because the autonomous zebrafish behavior recorded by \citet{mu2019glia} is primarily driven by visual input, we restrict the sensory stream to a vision encoder operating on images with similar resolution to the visual acuity of zebrafish \citep{haug2010visual}. Sensory features are used by three distinct cognitive networks with modularized objectives. The core and policy modules are implemented as an actor-critic architecture using Long Short-term Memory (LSTM) networks~\citep{hochreiter1997long} followed by feedforward decoders trained end-to-end with Proximal Policy Optimization (PPO) to output torques that are used as the control input to the agent's motors~\citep{schulman2017proximal}. Specific agent implementation  details can be found in Appendix \ref{sec:appendix-implementation-details}.
Although PPO is a model-free reinforcement learning algorithm, the intrinsic drive module (IDM) learns a world model through online experience that approximates the world state-action transition dynamics in sensory feature-space (as described below). However, this internal model functions only to generate intrinsic reward and is not used for planning or model-based control. 

\begin{figure}
    \centering
    \includegraphics[width=1\linewidth]{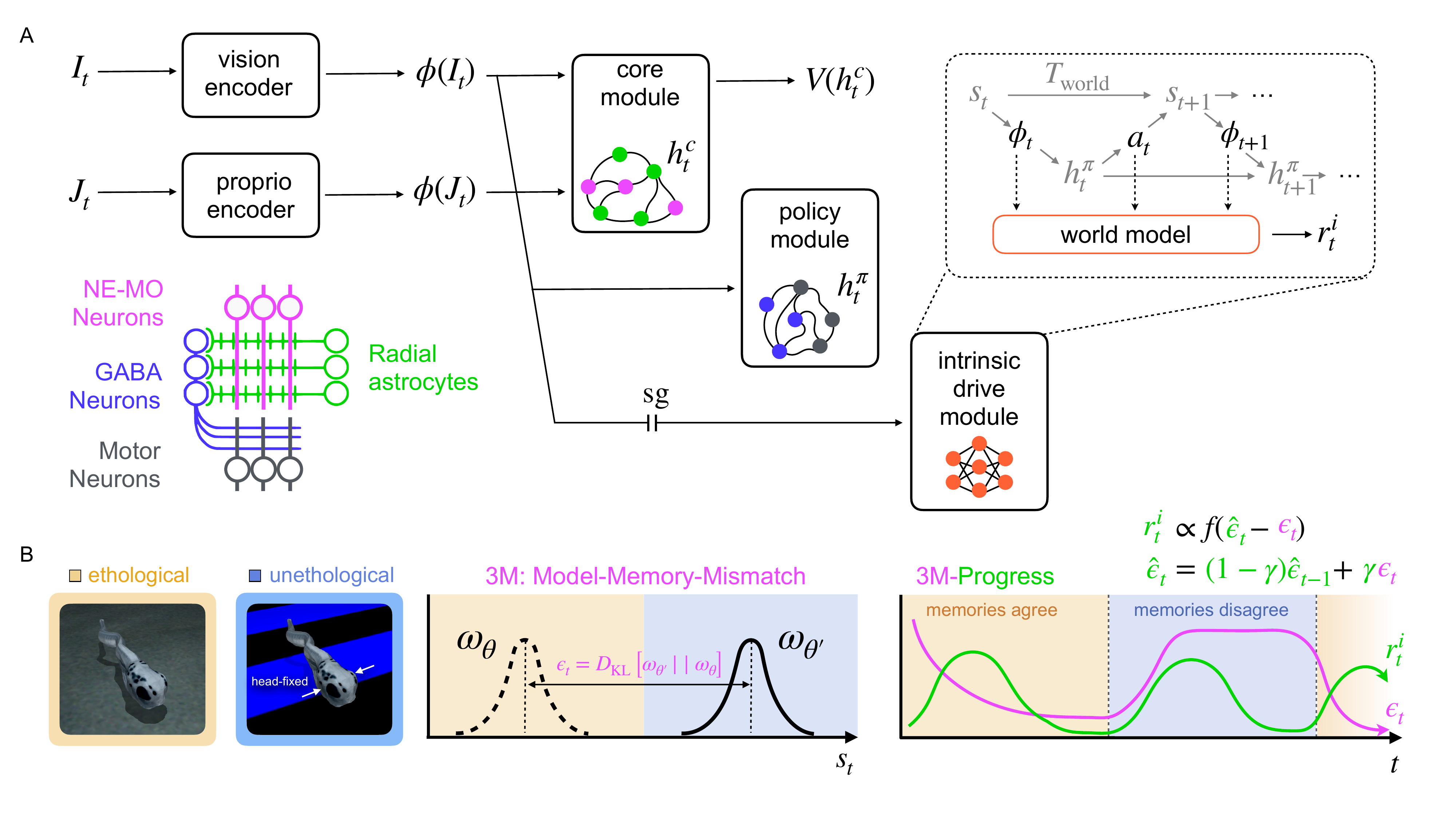}
    \caption{Agent architecture and 3M-Progress. A) Egocentric visual input ($I_t)$ is encoded via a small residual network $\phi_I$. Proprioceptive state observations ($J_t$) are encoded via a small multi-layer perceptron $\phi_J$ with shortcut paths to both the core and policy module. 
    Sensory features are passed into recurrent LSTM core ($h_t^c$) and policy ($h_t^\pi$) modules that learn a state value function and stochastic policy, respectively. The intrinsic drive module consists of a small multi-layer perceptron that parameterizes a forward dynamics model on sensory features observed from an environment with dynamics $T_{\mathrm{world}}$. B) 3M-Progress uses two memories created from environments with differing transition dynamics. Divergence between the ethological prior $\omega_\theta$ and the current world-model $\omega_{\theta'}$ defines 3M, which is then is used as input to leaky integrator $\hat\epsilon$ to generate intrinsic reward $r_t^i$.}
    \label{fig:model-schematic}
    \vspace{-0.1cm}
\end{figure}

\paragraph{Model-based Intrinsic Reinforcement Learning} In classical reinforcement learning, policies are obtained by mapping a desired behavior to maximizing a reward function~\citep{sutton1998reinforcement}. 
This function is one component of the Markov Decision Process (MDP) that formally specifies a task $M$ as a tuple $\left(\mathcal{S}, \mathcal{A}, T, p_0, r\right)$, where $\mathcal{S}$ is the space of environment states, $\mathcal{A}$ is the space of actions, $T:\mathcal{S}\times\mathcal{A}\rightarrow \mathcal{P}(\mathcal{S})$ is the transition dynamics (where $\mathcal{P}(\mathcal{S})$ is the set of probability densities over $\mathcal{S}$), $p_0$ is the distribution over initial states, and $r$ is the reward function. 
Importantly, when reward is defined as part of the task MDP, it is \textit{extrinsic} and provided from the environment. 
However, learning complex behaviors using extrinsic reward can fail for many real-world situations where $r$ is extremely sparse (winning a long-horizon game like Go) or intractable to begin with (intellectual pursuits such as knowledge acquisition) without a powerful exploration mechanism to guide behavior. 
Model-based intrinsic motivation is a class of such mechanisms that leverage predictive world models to convey exploration-relevant information using prediction-error to the form an intrinsic reward $r_t^i$. In the absence of any extrinsic reward $r_t^e$, policy learning with intrinsic motivation is completely self-supervised. 
In this work, we consider intrinsic motivation driven by forward dynamics world models $\omega$. 
Let $\theta$ parameterize a neural network and $\omega_\theta:\mathcal{S}\times\mathcal{A}\rightarrow\mathcal{P}(S)$. For simplicity, we assume a fixed variance Gaussian density $\omega_\theta=\mathcal{N}\left(\phi(\mathbf s_{t+1})\mid \phi(\mathbf s_t),\mathbf a_t; \mu_\theta, \sigma I\right)$. Here, $\phi(\cdot)$ denotes the concatenated sensory embeddings from $\phi_I(\cdot)$ and $\phi_J(\cdot)$. This class of methods is appealing from two complimentary perspectives. On one hand, it is computationally straightforward; learning an internal model of an MDP's transition dynamics is a natural way to measure novelty of the states visited under the agent's policy---states with a high prediction error under the internal model indicate regions of the state-action space that are poorly understood or rarely visited. On the other hand, it is well motivated by experimental evidence in both neuroscience and psychology; numerous empirical studies suggest humans and animals depend on predictive models of the world for decision-making~\citep{nejad2025self, daw2011model, gershman2018successor, bubic2010prediction, de2010predictive}. Thus, world models are well-positioned as a primary substrate for intrinsic motivation; however, we emphasize that a world model is best viewed as \textit{only a substrate}, as exactly how it should be used for intrinsic motivation remains unclear. 
\begin{figure}
    \centering
    \includegraphics[width=1\linewidth]{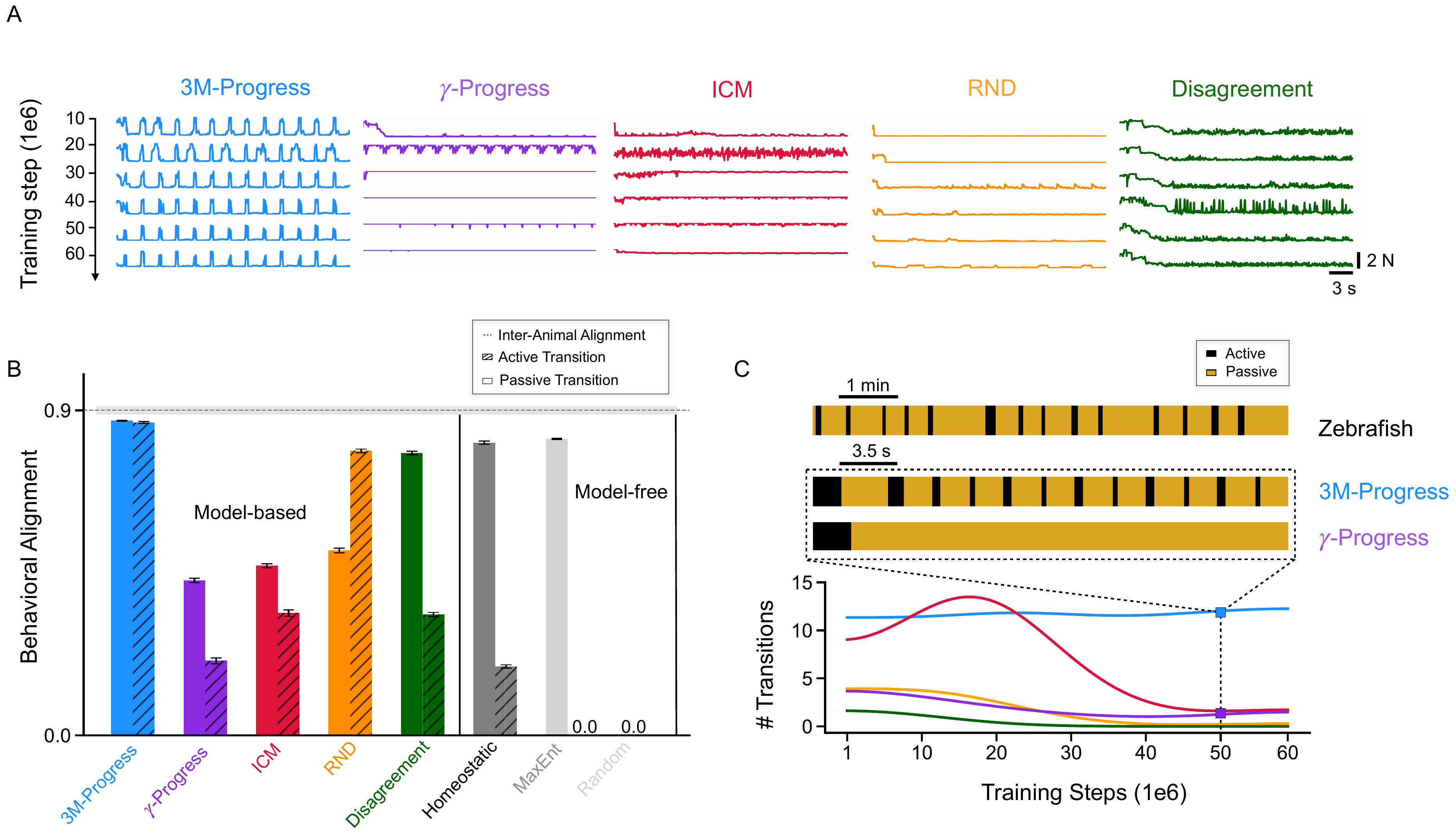}
    \caption{Model-behavioral alignment. A) Swim power traces of artificial agents with different intrinsic drives throughout training. B) Pearson's $r$ correlation between agent swim power (joint torques) and zebrafish swim power (motor neuron activity) for active and passive behavioral transitions. C) (Top) Timecourse of active-passive transitions in zebrafish compared against stationary behavior from progress-driven agents for a single rollout. (Bottom) Average number of behavioral transitions per rollout across training for different intrinsic drives.}
    \label{fig:behavior-alignment}
    \vspace{-5mm}

\end{figure}

\paragraph{Model-Brain and Inter-Animal Alignment} To rigorously evaluate our embodied agents against whole-brain data collected by~\citet{mu2019glia}, neural-glial recordings from zebrafish were first aligned to periods specifically capturing behavioral state transitions between active swimming and futility-induced passivity. 
Correspondingly, latent states and predicted neural-glial responses from our agents were aligned to these same experimentally defined epochs using event-triggered averaging around transition events. 
To quantify model-brain alignment, we employed a stringent ``One-to-One'' mapping, in which each neural or astrocytic unit from the zebrafish brain was matched directly to the single most correlated artificial unit from the virtual embodied agent. 
Although this mapping is typically too restrictive for capturing inter-animal alignment in heterogeneous neural populations for brain-region-specific (e.g. \emph{not} whole brain) data in other animals (as shown in prior sensory~\citep{nayebi2021unsupervised} and cognitive systems~\citep{nayebi2021explaining} where full linear regression is necessary), we found it was fully sufficient here due to the exceptional whole-brain reliability of zebrafish neural-glial responses during futility-induced passivity. 
Indeed, inter-animal alignment computed from pairwise correlations between individual zebrafish was nearly 100\%, establishing a robust empirical ceiling for evaluating model performance.

Inter-animal alignment was computed as the correlation of neural-glial response patterns across pairs of zebrafish within aligned epochs, thus formally defining the upper bound of predictivity achievable by any candidate model (the mathematical details of the chosen metric can be found in Appendix \ref{sec:methods-interanimal}). 
This approach ensures that successful computational models achieve neural predictivity indistinguishable from biological measurements, within the bounds set by natural biological variability.
Together, these procedures allowed us to precisely quantify model-brain alignment; achieving close correspondence between artificial and biological responses under this strict One-to-One mapping criterion provides strong evidence that our intrinsic motivation-driven embodied agents effectively capture the detailed neural-glial dynamics underlying autonomous behavioral transitions in zebrafish.

\paragraph{Model-Behavioral Alignment} To quantify similarity between biological zebrafish and artificial agents, we use Pearson's $r$ as a metric between the respective swim power readouts for each system surrounding state transitions. In zebrafish, swim power was calculated as the standard deviation over a 10ms window of recorded tail motor nerve signals following \citet{mu2019glia}. We use the same passive and active transitions windows identified by \citet{mu2019glia} to compute the inter-animal behavioral alignment across 11 subjects by applying the metric to swim power after smoothing and normalization over a 20 second window surrounding the transition time. For the artificial agents, we take the norm of their joint torques as swim power and identify behavioral transitions as high-frequency changes in swim power above a threshold of 1 Newton. This threshold was determined empirically by recording joint activations during active and passive behaviors in the default swim task in the dm-control suite 6-link swimmer environment \citep{tassa2018deepmind}. Model-behavioral consistency is then computed by applying the metric between segmented zebrafish data and the agent's swim power surrounding these transitions over a 20-step window. 

\section{Animal-like Exploration from First Principles: Lessons from Zebrafish}
Autonomy in animals enables intelligent and robust decision-making in complex environments, even in the presence of high-entropy or unfamiliar dynamics such as noisy stimuli or laboratory habitats. Reflecting on characteristics of autonomous exploration in zebrafish, we propose two simple desiderata for intrinsic drives that capture animal-like autonomy:
\begin{enumerate}
    \item \textit{Animals do not perseverate on unpredictable stimuli or pursue stimuli they cannot causally interact with}. \citet{mu2019glia} demonstrate that zebrafish transition to passive behavioral states  when motor commands elicit unpredictable sensory-feedback (unlearnable dynamics) or when sensory-feedback is withheld altogether (uncontrollable dynamics). An intrinsic objective for animal autonomy should avoid unpredictable or uncontrollable stimuli.
    \item \textit{Animals exhibit consistent decision-making strategies across repeated encounters of the same context}. Zebrafish exhibit stable behavioral state-switching in the open-loop experimental protocol across trials and subjects, ultimately converging on a single exploration policy \citep{mu2019glia}. An intrinsic objective for animal autonomy should converge to a stable behavioral policy.
\end{enumerate}

Intrinsic objectives that rely on prediction-error alone \citep{pathak2017curiosity, burda2018exploration} reward stochastic environment dynamics and incentivize learning transitions in which the agent has no causal control\footnote{Although \citep{pathak2017curiosity} proposed an inverse dynamics feature space to avoid representing uncontrollable stimuli, it's efficacy has not been demonstrated outside of simple pixel-based environments with discrete actions.}. Learning progress \citep{kim2020active, schmidhuber2010formal, achiam2017surprise} and Disagreement \citep{pathak2019self} overcome this by leveraging temporal dynamics or statistics of a world model ensemble, but together with ICM and RND, are formulated as functions of the world model training loss and are thus non-stationary---learning on repeated behavioral strategies drive the training loss towards zero, and so any single behavioral strategy is transient since it is not consistently reinforced. Although these properties are suitable for exploration in some robotics domains, particularly when the policy is supplemented with an extrinsic task reward \citep{sun2025curiosity, mendonca2021discovering, sekar2020planning, pathak2017curiosity, pathak2019self}, they fail to capture the nature of autonomous exploration in animals. 

\paragraph{3M-Progress} In order to overcome these drawbacks, we introduce Model-Memory-Mismatch Progress (3M-Progress), a novel intrinsic drive that incorporates these simple normative properties of animal autonomy inspired by zebrafish. Curiosity, disagreement, and learning progress couple intrinsic reward to a moving world model, yielding a non-stationary policy objective. The RL problem becomes a two player minimax game, where the actor seeks states that increase prediction-error and the learner reduces it, leading to a reward landscape that flattens and precludes a stationary optimal policy\footnote{See Appendix \ref{appendix: non-convergence} for mathematical details.}. Thus, each of the existing curiosity-driven exploration algorithms we consider fails to converge to an exploratory policy. However, learning progress \citep{kim2020active} is unique among these algorithms in its ability to avoid perseveration on unpredictable and uncontrollable stimuli by leveraging the temporal dynamics of prediction-error. This is achieved by maintaining two world models, an online model and a long-term memory implemented as an exponential weight decay on the online model. The slow timescale memory provides a moving baseline that, when compared with the online model, creates an intrinsic reward that flattens as the temporal dynamics of the online model's predictions flatten---in other words, a computationally efficient and biologically plausible implementation of a time-derivative. 

To combine the adaptive behavior afforded by derivative-like operations with an intrinsic goal that admits a stationary solution, we propose to  decouple the two memories completely. We achieve this by learning the long-term memory $\omega_\theta$ in a pretraining environment and deploying it as a fixed prior while the online model $\omega_{\theta'}$ is learned in a new environment with new transition dynamics. 
The frozen ``ethological memory'' functions as a static memory primitive: comparing its predictions with the online memory produces a residual that encodes an explicit bias toward transition dynamics that match the pretraining environment, which can then be leveraged to partition the state-action space in the new context into regions where memories systematically agree or disagree \footnote{See Appendix \ref{appendix: 3MP partitions} for mathematical details.}. This reflects the idea that animals develop in an ecological niche with characteristic environment dynamics, distilled via experience as an internal world model  $\omega_\theta$. In new environments with different transition dynamics, such as the experimental protocol in \citet{mu2019glia}, animals can use this prior to seek or avoid regions that match their ecological niche as they learn new world model $\omega_{\theta'}$. To capture this, we define the model-memory-mismatch ($\epsilon_t$), exponential filter ($\hat \epsilon_t$), and niche-aware intrinsic motivation $r_t^i$ as
\begin{align}
    \epsilon_t &:= \mathcal{D}_{\mathrm{KL}}\left[\omega_{\theta}\left(\phi({\mathbf s}_{t+1})\mid \phi({\mathbf s}_t), \mathbf a_t\right) \mid\mid \omega_{\theta'}\left(\phi({\mathbf s}_{t+1})\mid \phi({\mathbf s}_t), \mathbf a_t\right)\right],\\
    \hat \epsilon_t&=(1-\gamma)\hat\epsilon_{t-1}+\gamma\epsilon_t,\\
    r_t^i&=|\hat\epsilon_t-\epsilon_t|,
\end{align}
where $\mathcal{D}_{\mathrm{KL}}$ denotes the Kullback-Leibler divergence and $\gamma$ is the filter timescale. The non-negativity of this divergence coarsely partitions the state-action space into niche-seeking (model-memory agreement: $\epsilon_t\approx0$) and niche-avoidance (model-memory disagreement: $\epsilon_t\gg 0$). We design the intrinsic reward mechanistically after $\gamma$-Progress \citep{kim2020active} to maintain a moving baseline, but filter model-memory-mismatch rather than model parameters. 

\begin{figure}
    \centering
    \includegraphics[width=1\linewidth]{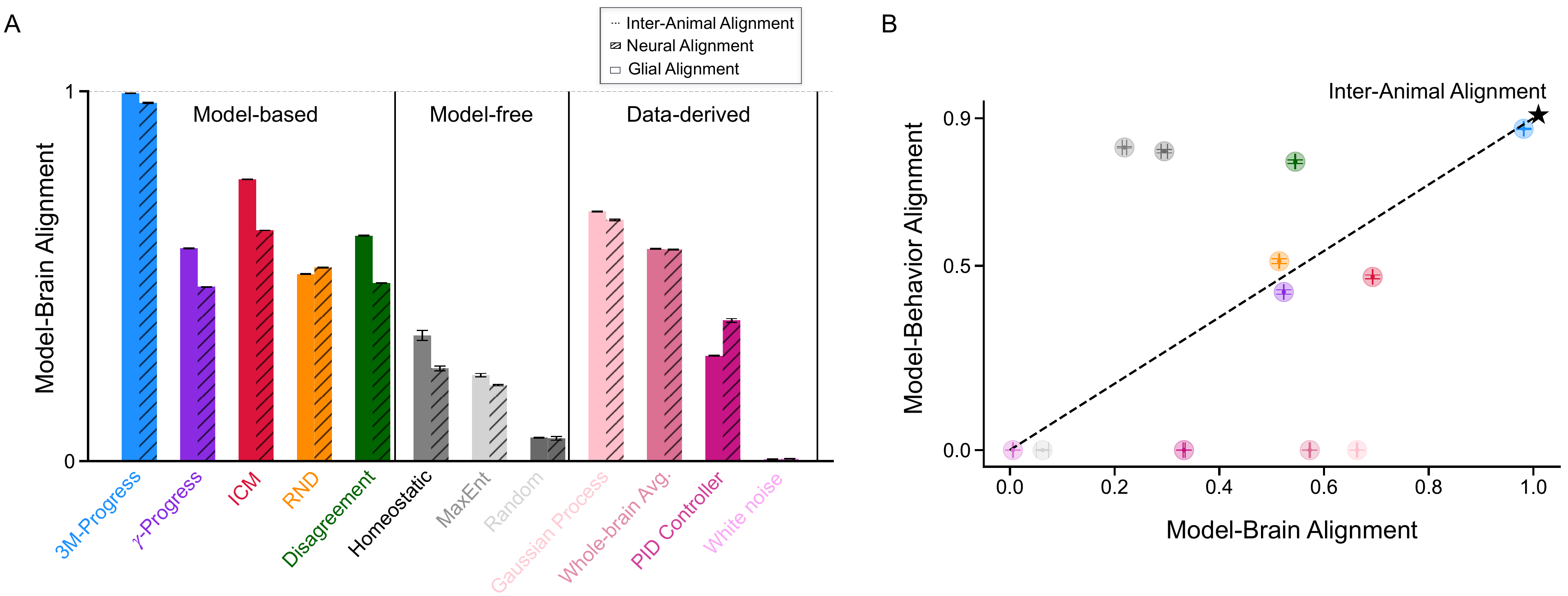}
    \caption{Model-brain alignment averaged across active and passive transitions. A) Noise-corrected Pearson's $r$ correlation between whole-brain neural and glial units and artificial units from trained agents. B) Model scores on behavioral and whole-brain alignment.}
    \label{fig:brain-model-alignment}
    \vspace{-0.1cm}
\end{figure}
Due to the moving baseline, the reward does not perseverate on any single partition, either agreement or disagreement. The symmetry enforced by the absolute value encourages periodic exploration between partitions by reinforcing deviations from the moving baseline in either direction. Unlike learning progress, this formulation does not saturate as learning unfolds since $\epsilon_t$ is computed using a fixed memory. In fact, as the prediction error from $\omega_{\theta'}$ stabilizes with more environment interactions, the signal-to-noise ratio in $\epsilon_t$ only increases, resulting in more robust behavioral patterns. Note that the absolute value is a specific choice of an activation function. The shape of this function and the progress horizon $\gamma$ determine the relative time spent in each partition. All experiments and baseline algorithms include an action penalty $r^a_t=-\lambda\Vert\mathbf a_t\Vert_2^2$ to encourage exploration of passive behavior. Figure \ref{fig:model-schematic}B illustrates a specific example inspired by zebrafish, where $\omega_\theta$ is learned in an ethological environment---the agent can freely behave and experiences passive fluid forces induced by self-motion as it swims and and active fluid forces as it learns positional homeostasis by resisting an opposing current. 
The agent is then put into an unethological experimental protocol where it is head-fixed and its swim commands do not elicit sensory feedback. As the agent behaves in this environment, it distills a new memory $\omega_{\theta'}$ from experience.

\paragraph{3M-Progress as a Normative Model of Neural-Glial Computation} Owing to the formulation of 3M-Progress, the state-value function must implement units in the core module that are functionally equivalent to neurons in the Noradernergic cluster of the Medulla Oblongata (NE-MO) and radial astrocytes identified by \citet{mu2019glia} (\ref{fig:model-schematic}A). 3M-Progress detects sensory-motor mismatch using a prior memory as an \textit{expectation} of how action is coupled to sensory-feedback under ethological environment dynamics (\ref{fig:model-schematic}B). This is functionally equivalent to signaling failed swim-attempts by NE-MO neurons in zebrafish.
Similarly, the exponential filter is a discrete-time leaky integrator on model-memory-mismatch (NE-MO input), which is functionally equivalent to radial astrocytes that accumulate NE-MO signals during failed swim-attempts and decay during passivity (\ref{fig:model-schematic}C). 



\section{Experimental Results}
\label{sec:results}

\paragraph{3M-Progress Agents Replicate Behavioral Patterns Observed in Zebrafish}
We first assessed the ability of intrinsic motivation methods to replicate detailed behavioral patterns observed in biological zebrafish during autonomous exploration. 
Behavioral alignment was quantified by comparing locomotor trajectories and state transitions (active to passive and back) between artificial agents and zebrafish across multiple trials and subjects. \begin{wrapfigure}{r}{0.55\linewidth}
    \vspace{-0.35cm}
    \centering
    \includegraphics[width=\linewidth]{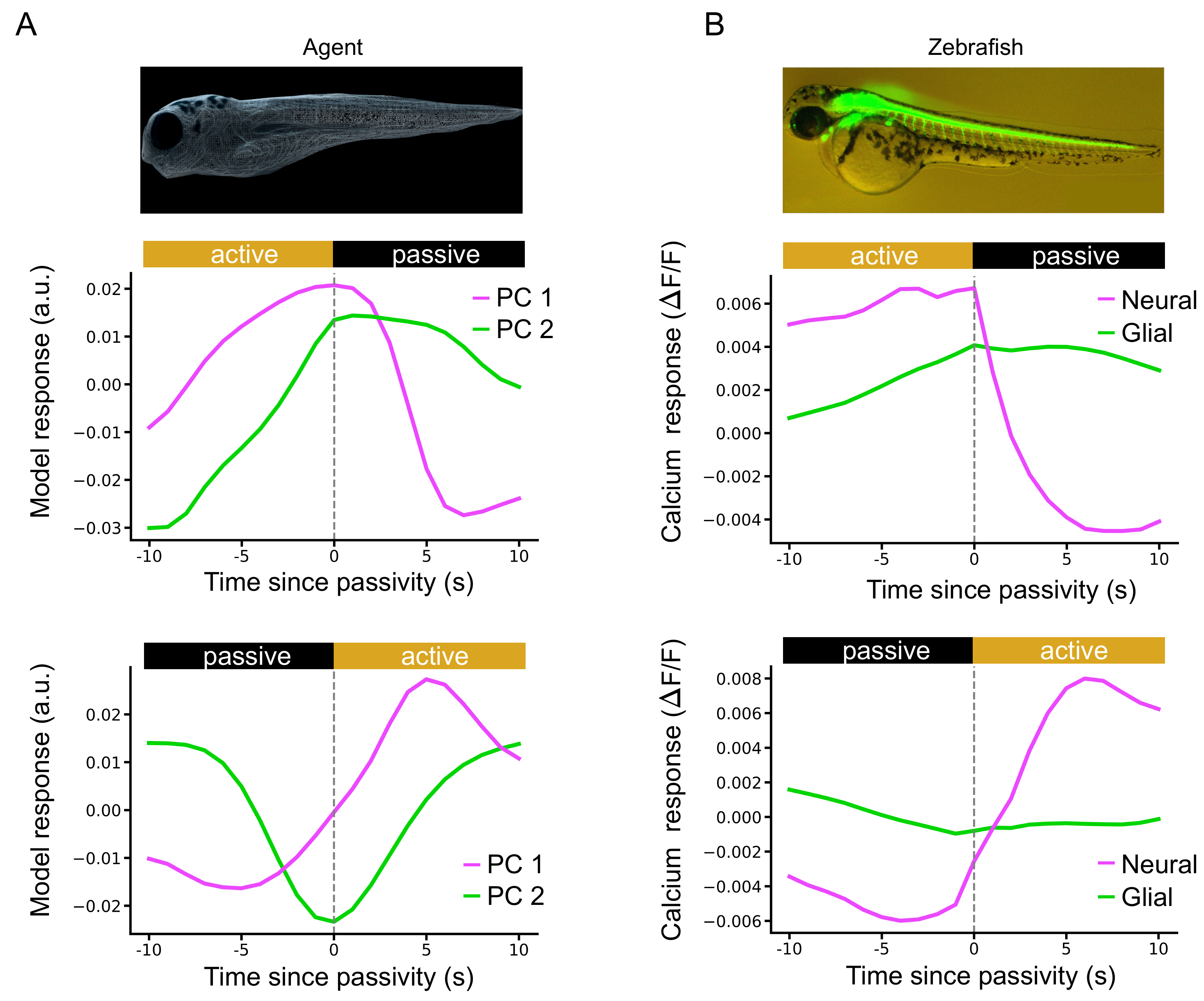}
    \caption{Latent dynamics of 3M-Progress agent's internal activations compared with normalized whole-brain neural-glial response in zebrafish. A) Principal components during passive and active transitions in the agent. B) Normalized average whole-brain neural-glial response during passive and active transitions in a zebrafish subject.}
    \label{fig:neural-glial-pca}
    \vspace{-1cm}
\end{wrapfigure} 
3M-Progress discovers ethologically-relevant behavioral state transitions by 10 million environment steps and sooner, whereas other intrinsic drives display transient strategies 
and stabilize on complete passivity or activity over training (Figure \ref{fig:behavior-alignment}A). 
Agents trained with 3M-Progress exhibited the highest model-behavior alignment, 
successfully capturing the dynamics of state transitions recorded in the biological data (Figure \ref{fig:behavior-alignment}B). Additionally, agents trained with traditional intrinsic motivation methods such as ICM, RND, Disagreement, and $\gamma$-progress showed significantly lower alignment and failed to capture the characteristic stable cycling between active and passive states (Figure \ref{fig:behavior-alignment}B-C). 

\paragraph{3M-Progress Agents Saturate Explainable Variance of Whole-Brain Neural-Glial Dynamics}

To evaluate how closely intrinsic motivation-driven agents matched zebrafish neural-glial dynamics, we compared model predictions to whole-brain calcium imaging data recorded during behavioral transitions of futility-induced passivity. 
Neural-glial alignment was quantified using a One-to-One mapping, where each recorded biological neuron and astrocyte was matched to the single most correlated artificial unit from the model. Strikingly, 3M-Progress agents captured nearly all of the explainable variance in neural and astrocytic activity, markedly outperforming existing intrinsic motivation algorithms, as well as baseline controls (Figure \ref{fig:brain-model-alignment}A). Model-free intrinsic drive controls include energy cost (homeostatic), entropy bonus (MaxEnt), and a randomly intialized agent. Data-derived controls include a Gaussian Process fit to neural-glial responses,  proportional-integral-derivative (PID) control, average neural-glial population response, and white-noise. More details on these controls can be found in Appendix \ref{sec:appendix-description of control models}. 
In fact, the 3M-Progress agent is the only model that is almost completely aligned with \emph{both} the behavioral alignment and neural-glial alignment, highlighting that accurate behavioral modeling can tightly constrain detailed neural dynamics (Figure \ref{fig:brain-model-alignment}B). Taken together, 3M-Progress agents pass the NeuroAI Turing Test on this dataset, a criterion emphasizing models that match both behavior and internal function~\citep{feather2025brain}.

\paragraph{Latent Dynamics of 3M-Progress Agents Reflect Underlying Neural-Glial Computations}

Given that 3M-Progress best matched both behavioral and neural-glial alignment among all candidate models, we then characterized the agent's internal dynamics by performing Principal Component Analysis (PCA) on its \emph{in silico} neural-glial population during behavioral state transitions. PCA revealed that the dominant latent dimensions of the agent's core module closely mirrors whole-brain neural-glial dynamics measured from biological zebrafish, whereby glial responses accumulate evidence of motor futility via noradrenergic signaling to drive behavioral suppression, and neural responses reflect transient activation patterns associated with detecting mismatches between expected and actual sensory outcomes during unsuccessful swim attempts. 
This analysis demonstrates that the 3M-Progress mechanism for intrinsic motivation not only generates realistic behavior, but also robustly captures fundamental internal neural-glial computations underlying autonomous exploration and behavioral state transitions (Figure \ref{fig:neural-glial-pca}).
\vspace{-0.05cm}
\section{Discussion}
\label{sec:discuss}

Our work seeks to identify and computationalize intrinsic goals that enable autonomous, task-independent behavior in animals. Leveraging a unique whole-brain dataset recorded in larval zebrafish during an autonomous behavior known as futility-induced passivity, we identify two simple principles of intrinsic goals for autonomous agents: avoid perseveration on stimuli that are uncontrollable and unpredictable, and converge on a robust decision-making strategy. We introduced 3M-Progress, a novel intrinsic drive that operationalizes these principles by continually tracking divergence between an online world model and an ethologically relevant prior. Learned from experience in an ecological niche that captures the basic physics of a naturalistic zebrafish habitat, this prior guides exploration in new environments by partitioning the behavioral space into niche-seeking and niche-avoiding modes. 

Unlike prior intrinsic motivation methods that suffer from behavioral inconsistency and non-stationarity, 3M-Progress reliably generated stable cycling behaviors closely matching those observed in biological zebrafish. Moreover, 3M-Progress agents were uniquely successful in capturing whole-brain neural-glial activity among all candidate models.
We showed that artificial agent's internal latent dynamics mirror neural-glial computation, wherein astrocytic responses accumulate evidence of motor futility through noradrenergic signaling to trigger behavioral suppression, while neural populations transiently encode mismatches between expected and observed sensory outcomes. To the best of our knowledge, our work marks the first goal-driven model of neural-glial computation, as well as the first completely self-supervised embodied agent that predicts behavioral and brain data. 

Our findings suggest two complementary evolutionary constraints for developing robust, animal-like autonomous agents: (1) maintaining an intrinsic drive informed by memory, and (2) continually monitoring divergence between this memory and new sensory experiences.
Functionally, tracking this mismatch within reinforcement learning frameworks allows artificial agents to identify ineffective strategies, update internal models adaptively, and discover new behaviors. 
Such mechanisms offer significant potential for enhancing the autonomy of artificial systems, especially in open-ended environments lacking clear external rewards or goals.
\vspace{-0.05cm}
\paragraph{Limitations and Future Work}
Our current analyses primarily focused on behavioral transitions within constrained virtual environments, limiting ecological realism. 
Future work could extend both the artificial environments and biological experiments to richer ecological settings and more complex behavioral repertoires, providing stronger and more diverse constraints on computational models of autonomy. The biomechanical realism of the body could also be expanded to include muscles and motor circuits, providing realistic constraints on the low-level controller.
Additionally, while our model captures essential neuron-glial interactions at a population level, it abstracts away detailed biochemical signaling mechanisms and anatomy of astrocytes and neurons. Incorporating more biologically detailed models of these processes could aid in providing predictions about these mechanisms at a finer scale. Finally, 3M-Progress can be generalized in a variety of ways that extend beyond futility-induced passivity (see Appendix \ref{appendix: 3MP generalization})---we leave this to future work. 
\newpage
\section*{Acknowledgements}
We thank Misha Ahrens, Chris Doyle, and Yu Mu for helpful discussions, as well as the anonymous reviewers for their helpful feedback on the initial manuscript draft.
A.N. was supported in part by a grant from the Burroughs Wellcome Fund. X.P. and R.K. were supported in part by the National Science Foundation and DoD OUSD (R \& E) under Cooperative Agreement PHY-2229929 (The NSF AI Institute for Artificial and Natural Intelligence, ARNI) and the Simons Collaboration in Ecological Neuroscience (SFI-AN-NC-SCN-00007276-10).
\bibliography{refs}
\bibliographystyle{unsrtnat}
\newpage
\section*{Appendix}
\appendix
\section{Model-Brain Alignment per Transition and per Module}
\label{sec:appendix-alignment-per-transition-per-module}

\begin{figure}[!htbp]
  \centering
  \includegraphics[width=\linewidth]{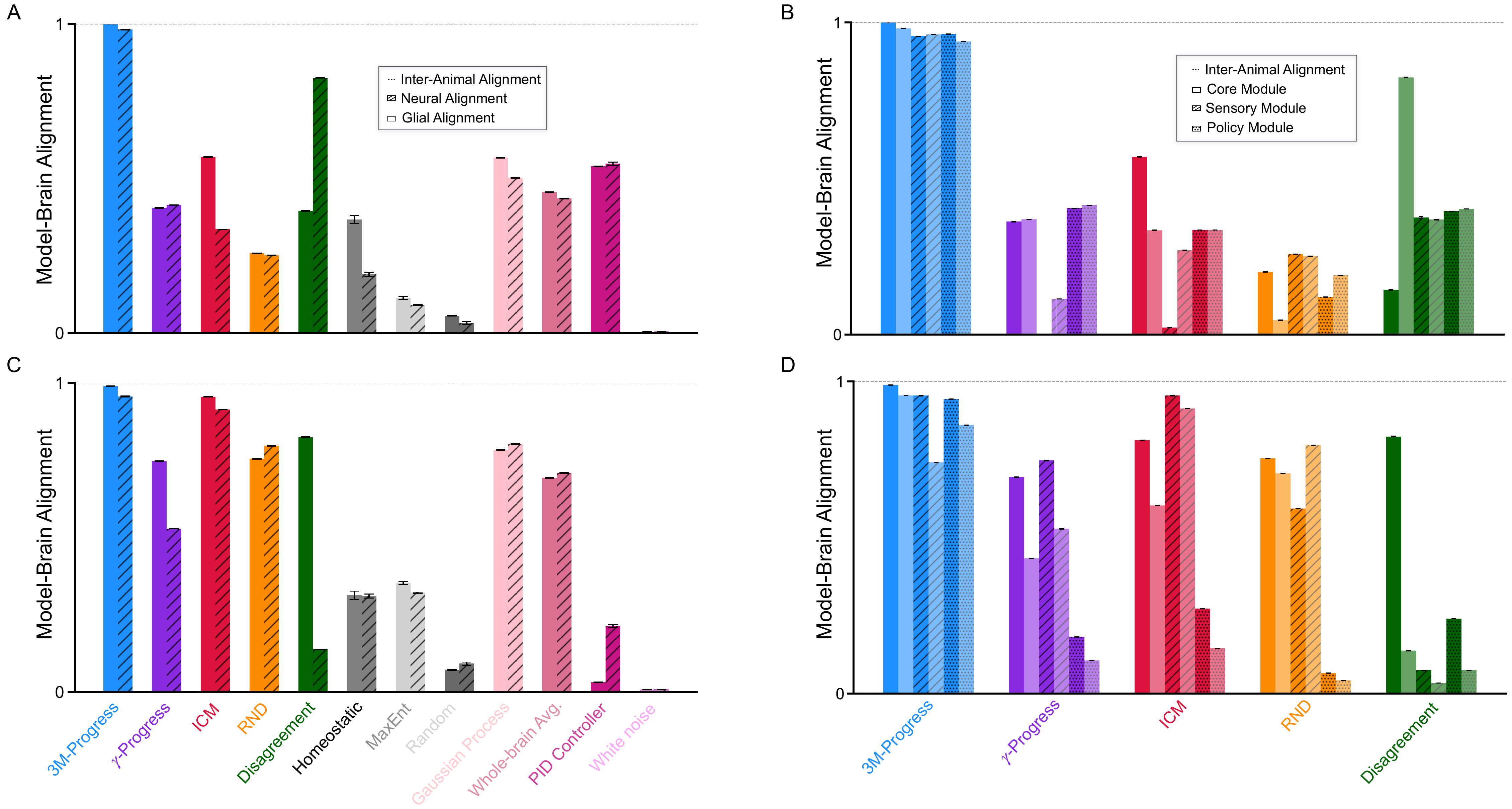}
  \caption{Model-Brain alignment for active and passive transitions and per module, excluding per module read-in and readout layers where applicable. A) Alignment for Active Transitions. B) Alignment per Agent Module for Active Transitions. C) Alignment for Passive Transitions. D) Alignment per Agent Network Module for Passive Transitions.}
  \label{fig:model-brain-alignment-per-transition-per-model}
\end{figure}
\section{Latent Dynamics of Baseline Agents}

\begin{figure}[!htbp]
    \centering
    \includegraphics[width=1\linewidth]{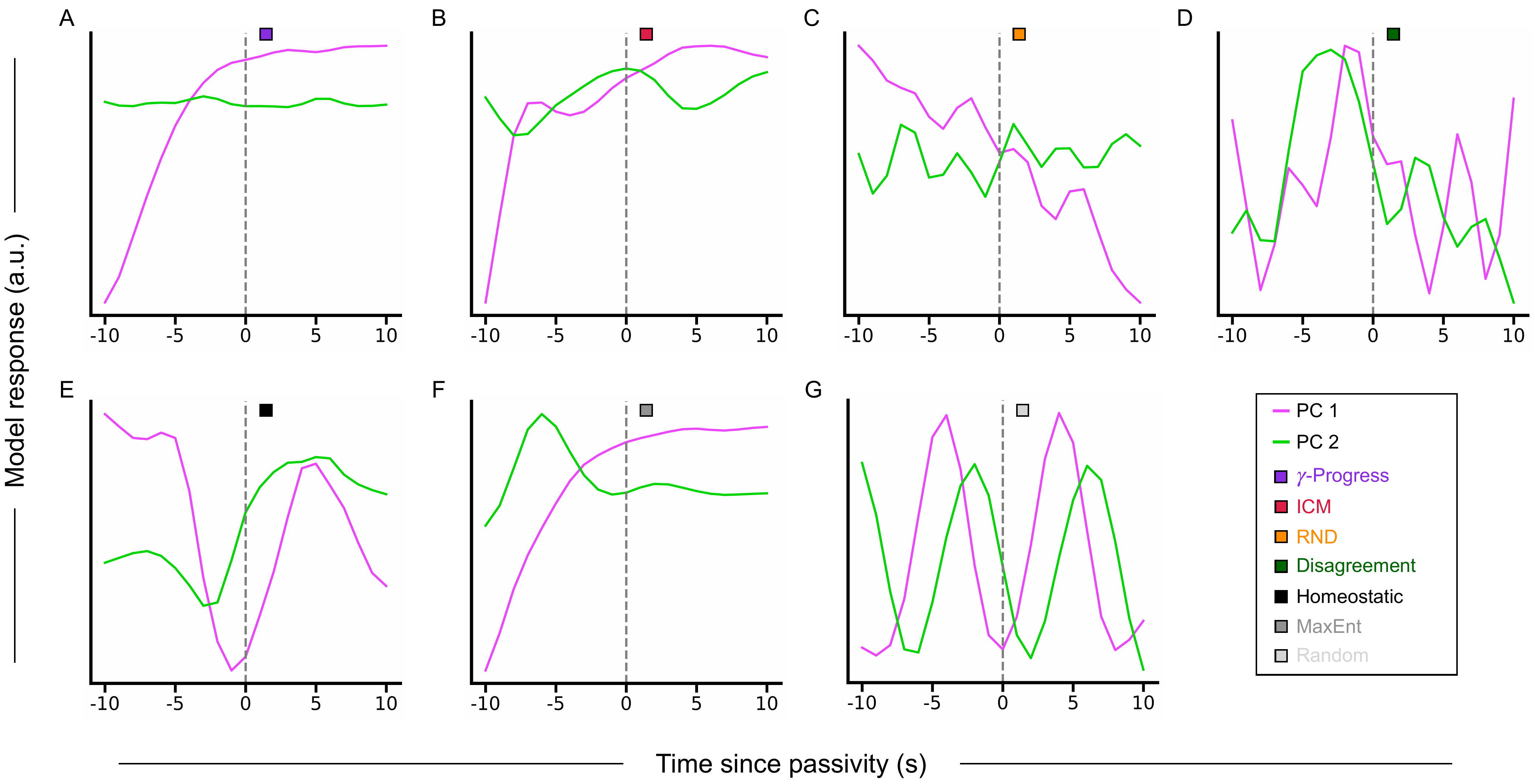}
    \caption{Latent dynamics of each agent-based control model. Dashed line indicates time since passivity. Coloring of PCs neural-glial cell-types (Fig. \ref{fig:model-schematic}, Fig. \ref{fig:neural-glial-pca}) was chosen according to cell-types of the 3M-Progress agent PCs.}
    \label{fig:placeholder}
\end{figure}
\section{Formal Intuitions}
\subsection{Curiosity-driven Exploration is Not Enough} \label{appendix: non-convergence}
Consider an MDP with discount $\gamma\in(0,1)$, occupancy $d_{\pi}(s,a)=(1-\gamma)\sum_{t\ge 0}\gamma^t\Pr_{\pi}(s_t{=}s,a_t{=}a)$, and transition kernel $\mathcal{T}(s'\mid s, a)$. Let a predictive world model with parameters $\theta$ be trained online to minimize
\[
\mathcal{L}(\pi,\theta)\;=\;\mathbb{E}_{(s,a,s')\sim d_\pi \mathcal T}\big[\ell(s,a,s';\theta)\big],
\]
and suppose the policy receives intrinsic reward determined by this predictor: $r^{i}(s,a,s';\theta)\;=\;g\big(\ell(s,a,s';\theta)\big)$
where $\ell\ge 0$ and $g:\mathbb{R}_{\ge 0}\to\mathbb{R}_{\ge 0}$ is monotone increasing with $g(0)=0$.  The policy is updated to  maximize $$J(\pi,\theta)\;=\;\mathbb{E}_{(s,a,s')\sim d_\pi \mathcal T}\!\big[r^{i}(s,a,s';\theta)\big].$$ Under either alternating or simultaneous updates that decrease $\mathcal{L}$ in $\theta$ and increase $J$ in $\pi$, the process cannot converge to a nontrivial, uniquely defined exploratory policy. The only stationary outcome is a degenerate collapse in which $r^{i}$ is constant on the visited support; otherwise the policy-model coupling remains non-stationary and induces drift. That is, any stationary point $(\pi^\star, \theta^\star)$ satisfies one of the following cases:
\begin{enumerate}
\item \textit{Reward collapse.} If $\theta^\star$ minimizes $\mathcal{L}(\pi^\star,\theta)$ on the support of $d_{\pi^\star}\mathcal T$ and the model class is realizable on that support, then $\ell(\cdot;\theta^\star)=0$ almost surely, hence $r^{i}(\cdot;\theta^\star)=0$ almost surely and $J(\pi,\theta^\star)= 0$ for all $\pi$. The objective is flat and does {not} select a unique exploratory policy.
\item \textit{No stationary policy.}
If we freeze the model at $\theta^\star$, then $r^{i}(\cdot;\theta^\star)$ is a fixed reward. If residual error remains, then $\ell(\cdot;\theta^\star)$ (hence $r^{{i}}$) is not almost surely constant, so there exists a measurable set $U$ on which the intrinsic advantage $A^{r^{\mathrm{int}}}_{\pi^\star}(s,a)>0$ for some actions. By the policy-gradient identity,
\[
\nabla J(\pi^\star,\theta^\star) \;=\; (1-\gamma)\,\mathbb{E}_{d_{\pi^\star}}\!\big[A_{\pi^\star}(s,a)\,\nabla \log \pi^\star(a\mid s)\big].
\]
For some state $s\in U$, suppose we increase $\pi^\star(a\mid s)$ slightly on an action with $A^{}_{\pi^\star}(s,a)>0$ and decrease it on other actions to preserve $\sum_a\pi^\star(a\mid s)=1$. This choice makes the inner product $\langle A^{}_{\pi^\star}(s,\cdot), \nabla \log \pi^\star(\cdot\mid s)\rangle$ strictly positive on $U$, hence the expectation above is positive and the policy gradient is nonzero. Therefore $\pi^\star$ is not a local maximizer of the stationary objective. Any such occupancy shift then triggers predictor updates that reduce $r^{\mathrm{int}}$ on $U$, moving the high-reward region and preventing a nontrivial fixed point.

\end{enumerate}
In both cases, these intrinsic signals do not converge to a stable, uniquely-defined exploratory policy. Either training drives the intrinsic signal to a constant (degenerate) value on the visited distribution, or residual heterogeneity in the reward keeps creating ascent directions that are then neutralized by predictor updates, preventing stabilization. 
\subsection{3M-Progress Partitions the Behavioral Space} \label{appendix: 3MP partitions}
Consider two reward-free MDPs, $\mathcal{M}_1=\left({\mathcal{S}}, {\mathcal{A}}, \mathcal{T}_1, {p}_0\right)$ and ${\mathcal{M}}_2=\left({\mathcal{S}}, {\mathcal{A}}, {\mathcal T}_2, {p}_0\right)$, that differ only in their transition densities. Further, suppose there exists a set $U\subset \mathcal S\times \mathcal A$  such that $\mathcal T_1(\cdot\mid s, a)=\mathcal T_2(\cdot\mid s, a)$ almost everywhere in $s'$ for all $(s, a)\in U$. Let $p(\cdot\mid s, a)$ and $q(\cdot\mid s, a)$ be world models trained on data from $\mathcal T_1$ and $\mathcal T_2$, respectively, with pointwise consistency: 
$$\mathcal{D}_{\mathrm{KL}}\left(\mathcal T_1(\cdot\mid s, a) \|p(\cdot\mid s, a)\right)\overset{p}{\rightarrow} 0, \quad \mathcal{D}_{\mathrm{KL}}\left(\mathcal T_2(\cdot\mid s, a) \|q(\cdot\mid s, a)\right)\overset{p}{\rightarrow} 0.$$ Assume for each $(s, a)$, both kernels share the same support with no vanishing probabilities (i.e. $\exists c>0$ s.t. $\mathcal T_2(s'\mid s, a)\geq c$ whenever $\mathcal T_1(s'\mid s, a)>0)$. Then it follows that for all $(s, a)\in U$, 
$$\mathcal{D}_{\mathrm{KL}}\left(p(\cdot\mid s, a) \|q(\cdot\mid s, a)\right)\overset{p}{\rightarrow} 0,$$
and that for all $(s, a)\in U^c$, $$\mathcal{D}_{\mathrm{KL}}\left( p(\cdot\mid s, a) \|q(\cdot\mid s, a)\right)\overset{p}{\rightarrow}\mathcal{D}_{\mathrm{KL}}(\mathcal T_1\| \mathcal T_2)>0.$$

\subsection{Beyond Futility-induced Passivity} \label{appendix: 3MP generalization}

Although we demonstrate 3M-Progress on a specific autonomous behavior known as futility-induced passivity, our algorithm applies to any exploration behavior in which a dynamics niche is reasonably specified. In our experiments, futility-induced passivity arises completely from the choice of the pretraining environment and the online learning environment; we choose these environments such that the transition dynamics between environments agree when the agent is passive, thus defining the ecological niche that guides exploration (see Appendix \ref{appendix: 3MP partitions}: simply put, we choose environments such that there exists a subset $U$ in which the transition dynamics locally agree). In general, the pretraining environment should encode a meaningful dynamics prior the agent can use for continual learning with environments whose physics vary systematically from the pretraining environment.  

For example, suppose we pretrain an agent and world model on a foraging task (e.g. the virtual rodent environment in \citep{tunyasuvunakool2020dm_control}). In a new environment that includes the opportunity for foraging and any of its constituent locomotor primitives (running, turning, jumping, etc.), the basic distribution-matching objective of 3M-Progress rewards the agent for trajectories whose dynamics are predictable under the pretrained world model (namely, foraging and constituent motor primitives): 
$$r_t^i=f(\hat \epsilon_t-\epsilon_t); \quad \epsilon_t=\mathcal{D}_{\mathrm{KL}}(\omega_{\theta'}\|\omega_\theta); \quad \hat \epsilon_t = (1-\gamma)\hat\epsilon_t+\gamma\epsilon_t$$
where $\theta'$ is learned online and $\theta$ is pretrained. Niche-aware exploration is primarily mediated by the activation function $f$. When $f$ is monotonic non-decreasing, such as a rectified linear unit, 3M-Progress is niche-seeking with the niche defined by the dynamics prior $\omega_\theta$. Conversely, if $f$ is monotonic non-increasing, 3M-Progress is niche-avoiding and explores dynamics outside the prior. Non-monotonic functions allow some amount of symmetry between niche-seeking and niche-avoidance depending on the specific function shape. 

The utility of this approach can also be appreciated when the pretraining stage involves a large diversity of dynamics, either from multiple environments and tasks or a single environment with multiple tasks. A world model with sufficient computational capacity that captures these diverse dynamics can be flexibly used in new environments to motivate exploration in a variety of ways. In the simplest case, for example, suppose we pretrain an ensemble of dynamics models $\{\omega_{\theta_j}\}_{j=1}^N$ on $N$ separate environments and tasks. Maintaining independent temporal filters $\hat \epsilon_t^j$ for each prior, a deterministic intrinsic motivation can be defined as $r_t^i=\max \{f(\hat\epsilon_t^j-\epsilon_t^j)\}_{j=1}^N$. Alternatively, one can imagine various sampling schemes over the ensemble in order to drive specific exploration styles, such as $\epsilon$-greedy or max-entropy. This extends 3M-Progress to niche-aware exploration over multiple niches, thereby allowing the flexibility of multiple modes of behavior in a single objective function. Each addition of a dynamics prior further partitions the behavioral space in the online environment, and allows the dynamics characteristic of each pretraining environment to be composed to form an exploration landscape of attractors (niche-seeking) or repellors (niche-avoidance) for online learning. 
\section{Description of Control Models}
\label{sec:appendix-description of control models}
\subsection{Model-based Controls}
\paragraph{The Intrinsic Curiosity Module (ICM) }\citep{pathak2017curiosity} defines intrinsic reward as the Shannon surprise of the forward model, $r_t^i:=-\log\omega_\theta(\phi^I_t\mid \phi_t, \mathbf a_t)$. $I$ denotes an augmented inverse feature space that is learned on-top of sensory embeddings from $\phi$---ICM trains an additional embedding layer $\phi^I$ using an inverse dynamics model parameterized by $\theta_I$ by optimizing an MSE loss $L(\theta_I)=\mathbb{E}_{\pi_\theta} \left\Vert f(\theta_I; \phi^I_t, \phi^I_{t+1})-\mathbf a_t \right\Vert_2^2$.

\paragraph{Random Network Distillation (RND)} \citep{burda2018exploration} defines a fixed random nonlinear projection of sensory features  $g(\phi)$ and trains a predictor network $\hat{g}$ using an MSE loss $r_t^i:=L(\theta_{\mathrm{RND}})=\mathbb{E}_{\pi_\theta} \left\Vert g(\phi_t)-\hat{g}(\theta_{\mathrm{RND}}; \phi_t) \right\Vert_2^2$. With the distillation objective as the intrinsic reward, RND does not reinforce behaviors by scoring their predictability by a forward dynamics model as in ICM; instead, the random memory provides a simple exploration bonus for visiting novel states under the policy distribution.

\paragraph{Disagreement} \citep{pathak2019self} learn an ensemble of world models $\{\omega_{\theta_j}\}_{j=1}^N$ and defines intrinsic reward as $r_i^t:=\mathrm{Var}\left(\{\mu_{\theta_j}:j\in[N]\}\right)$ for $N$ randomly initialized world models. Ensemble variances scores the stochasticity of the environment and reinforces state-action pairs for which models disagree.

\paragraph{$\gamma$-Progress} \citep{schmidhuber2010formal} leverages the temporal history of Shannon surprise to define intrinsic reward using prediction gain, $r_i^t:=
\log\frac{\omega_{\theta_{new}}}{{\omega_{\theta_{old}}}}$, where $\theta_{new}$ parameterized a world model after learning on new transitions withheld from an lagging model $\theta_{old}$.
\subsection{Model-Free Controls}
\paragraph{Homeostatic Agent} One straightforward way to achieve a passive behavioral transition is to simply add an action-cost that outweighs any other positive reward signal. In the presence of a fixed or nonexistent extrinsic reward signal which the agent has no control over (such as in the open-loop protocol), an action-cost encourages the agent to become passive, corresponding to a metabolic constraint or homeostatic regulation of energy. We implement this cost as the magnitude of the force exerted by the agent's motors, $c(a_t) = \lambda\lVert\mathbf a_t\rVert_2$. In all our experiments, we set $\lambda=1$. 

\paragraph{Maximum Entropy Agent} Maximum entropy RL is a general exploration strategy that provides a bonus reward proportional to the entropy of the current policy. That is, $r_t^i=\lambda\mathcal{H}\left[\pi(\mathbf a_t\mid \mathbf s_t)\right]$. In all our experiments, we set $\lambda=1$. 

\paragraph{Random Agent} To test a random baseline model for embodied control, we use a randomly initialized agent with the same model architecture (described in section $\ref{sec:methods}$, Figure \ref{fig:model-schematic}A, and in PPO implementation details below). 

\paragraph{Whole-brain Average} We use the average recorded neural-glial activity during a specified behavioral transition in one larval zebrafish to predict whole-brain neural-glial activity recorded from another larval zebrafish undergoing the same transition. The alignment under the metric described in section \ref{sec:methods} is computed using the average response of each cell-type (neural and glial) from the source animal to predict the corresponding cell-type in the target animal. We use two subjects and report the total alignment as the sample-weighted average between scores from both source-target pairs. 

\paragraph{Gaussian-Process} We fit a separate Gaussian Process (GP) to each cell-type (neural and glial) for each subject, using a radial basis function (RBF) kernel and centering the prior mean at the average whole-brain response. Alignment under the metric described in section \ref{sec:methods} is computed between the whole-brain data from the target cell-type from an individual subject and its corresponding GP as the source. We use two subjects and report the total alignment as the sample-weighted average between scores from both source-target pairs. 

\paragraph{White-noise} At each timestep $t$, the model's predicted next state is give by $x_{t+1}=x_t+\eta_t$, where $\eta_t\sim\mathcal{N}(0, 1)$ is white-noise. Because the metric that saturates inter-animal alignment is correlation-based, the mean and variance of this random walk are arbitrary. 

\paragraph{PID Controller}
To implement the circuit-based word model proposed by \citet{mu2019glia} for the zebrafish brain's transition from active to passive states, we employ a threshold-based (``bang--bang'') controller that switches the fish's swim power $P(t)$ between an active waveform $P_{\text{base}}(t)$ and complete cessation ($P(t) = 0$) once a cumulative ``futility'' signal exceeds a fixed GABAergic threshold. This controller can be interpreted as the high-gain limit of a saturated PID controller.

Perceived hydrostatic drift is modeled as a constant $v_d$. The motor plant converts swim power into a counter-drift locomotor velocity with gain $g_{\text{MS}}$,
\begin{equation*}
    v_s(t) = v_d - g_{\text{MS}} P(t).
\end{equation*}

Visual mismatch is the product of forward stimulus velocity and ongoing motor drive but is rectified to ignore overshoot,
\begin{equation*}
e(t) =
\begin{cases}
v_s(t) P(t), & \text{if } v_s(t) > 0, \\\\
0, & \text{otherwise}.
\end{cases}
\end{equation*}

A leaky integrator with time constant $\tau_F$ accumulates the mismatch (i.e., the futility):
\begin{equation*}
\dot{F}(t) = -\lambda_F F(t) + k_F e(t),
\end{equation*}
where $\lambda_F = 1/\tau_F$ and $k_F$ is a gain. A second leaky integrator converts sustained futility into an inhibitory drive,
\begin{align*}
\dot{G}(t) &= -\lambda_G G(t) + k_G \max(0, F(t) - \theta_F), \label{eq:gaba_drive} \\\\
P(t) &=
\begin{cases}
P_{\text{base}}(t), & G(t) \le \theta_G, \\\\
0, & G(t) > \theta_G.
\end{cases}
\end{align*}

To mimic experimental perturbations we set $g_{\text{MS}} = 0$ between $t_{\text{OL on}}$ and $t_{\text{OL off}}$, effectively clamping optic-flow feedback.



\section{Implementation Details}
All code can be found in \url{https://github.com/neuroagents-lab/autonomous\_zebrafish}.

\label{sec:appendix-implementation-details}
\paragraph{Proximal Policy Optimization (PPO)}
In all our experiments, we train our agents with PPO using a clipped surrogate objective \citep{schulman2017proximal} with $\epsilon^{CLIP}=0.2$ and normalized advantage function computed using Generalized Advantage Estimation (GAE) \citep{schulman2015high} with $\lambda^{GAE}=0.95$ . The policy network is an MLP with two hidden layers [128, 64] optimized by Adam \citep{diederik2014adam} with learning rate $\alpha=0.0003$ and gradients computed over 5 epochs of 1000-step trajectories with a 250-step batch size vectorized across 64 environments. This MLP parameterizes the policy as a 5-dimensional diagonal Gaussian distribution which is sampled to produce actions during training. For evaluation (the experiments in section \ref{sec:results}), the policy is deterministic by taking actions as the mean of the distribution. Actions take the form of continuous real-valued torques on $[-1, 1]$. The intrinsic rewards are normalized by dividing by a running estimate of the standard deviation of the sum of discounted rewards with discount factor $\gamma=0.99$, which then supervises the value network MLP with two hidden layers [128, 64] to estimate the expected discounted return using this same discount factor. Both policy and value networks compute on hidden states from separate LSTM blocks (Figure \ref{fig:model-schematic}A) using a shared embedding from the sensory feature extractors. Image embeddings are obtained from 64x64 pixel observations passed through a three-layer ResNet resulting in outputs with a spatial resolution of 16x16 (closely matching the visual acuity of larval zebrafish at the final layer). Image embeddings were flattened and concatenated with proprioceptive features including joint positions and rotational velocities and their embeddings from a two-layer MLP with hidden size [64, 64]. 

\paragraph{Intrinsic Drive Module (IDM)}
The IDM architecture details vary depending on which intrinsic drive it implements, and is described on a case-by-case basis in the sections below. Here, we describe the commonalities between intrinsic drives, which include the optimizer, forward dynamics loss, general forward model architecture, and memory buffer parameters. Each forward dynamics model across intrinsic drives is implemented as a two-layer MLP with hidden sizes [512, 512], trained to predict the true observation one time-step into the future from the current observation using an MSE loss (where inference is done in feature-space), and optimized using Adam \citep{diederik2014adam} with learning rate $\alpha=0.001$. The IDM maintains a memory buffer of the last 100 observation embeddings from each environment in the vector and trains it's constituent networks on this buffer every 20 steps. The IDM is trained for $1e^5$ steps before the intrinsic rewards are observed by the agent. 

\paragraph{3M-Progress} The ethological memory is created (with the default configuration outlined above) by training a swimmer agent on a simple navigation task in a head-free version of the experimental protocol outlined in section \ref{sec:methods}. 
This environment is chosen to provide the forward model with a similar visual input space as the head-fixed version while maintaining the ethology of unconstrained swimming in which the agent experiences naturalistic fluid forces and positional displacement in response to swim commands. The task is implemented as shaped reward proportional to the distance of the agent from a target location that in front of the agent, such that the swim-to-target behavior results in stabilizing the constant backwards flow of the high-contrast grating. The episode is long enough that optimizing this reward allows the agent to become passive was the target is reached. Together, our setup allows the agent to experience state-action-state triplets (current state, current action, and resulting state) associated with active and passive behaviors. This provides a close correspondence between sensory-motor coupling in our virtual ethological environment and the closed-loop experimental condition in \citet{mu2019glia}, where larval zebrafish swim against the passive backwards flow of high-contrast gratings motivated by positional homeostasis. Although our agent is motivated to swim by a different signal than its biological twin (i.e., moving towards a target location rather than a homeostatic drive that resists displacement from an opposing current), the design of the virtual environment renders the sensory-motor stream experienced by the internal world model equivalent between scenarios, since in both cases optic flow is counteracted by forward swim motion. Because this sensory-motor stream and it's resulting world-model alone define the ethological memory, the discrepancy in the signal that drove behavior has no bearing on training the new policy and value network in the open-loop environment (Figures \ref{fig:body-schematic}D, \ref{fig:model-schematic}B). 

In the open-loop environment, the agent randomly initializes a new world-model memory that is trained online with the default configuration outlined in the IDM section. The ethological world-model memory is loaded from earliest checkpoint where the agent achieved optimal swim-to-target behavior and it's weights are frozen. The model-memory-mismatch is computed as the MSE between the predictions from each memory, filtered by an exponential moving average with timescale $\gamma=0.99$, and the difference filtered and unfiltered predictions are passed through an $L_1$ activation. 
\paragraph{Random Network Distillation (RND) \citep{burda2018exploration}} For RND  both target and predictor networks are implemented using the default configuration in the IDM section. The predictor network is trained to predict random feature projections from the target as described in section $\ref{sec:methods}$. 

\paragraph{Intrinsic Curiosity Module (ICM) \citep{pathak2017curiosity}} In addition to a forward model implemented using the default configuration in the IDM section, the ICM implements an inverse dynamics model as an MLP with an identical architecture and optimization routine to train a one-layer MLP on top of sensory features. The forward and inverse networks are cotrained by minimizing a joint objective $\beta L_F + (1-\beta)L_I$, where $L_F$ and $L_I$ are the forward and inverse MSE loss functions, respectively. In our experiments, we set $\beta=0.2$. 
\paragraph{Disagreement \citep{pathak2019self}} We use $N=3$ randomly initialized independent forward models using the default configuration described in the IDM section. Disagreement is computed as the mean variance across feature dimensions. 
\paragraph{$\gamma$-Progress \citep{kim2020active}} Using a randomly initialized forward model implemented using the default configuration in the IDM section, the trailing memory is created by copying these initial weights and updating them using an exponential moving average with timescale $\gamma$. In all our experiments, we use $\gamma=0.99$. 

\section{Inter-Subject Noise Correction Derivation}
\label{sec:methods-interanimal}
Herein we describe how the metric $\mathcal{M}$ should correct for noise if there is trial-to-trial variability.
This is unified and adapted from~\citet{nayebi2021unsupervised,nayebi2021explaining,nayebi2023neural}.
If you prefer to skip the derivation, for common choices of metric $\mathcal{M}$, such as Pearson correlation, RSA, and especially any metric that satisfies transitive closure~\citep{williams2021generalized}, one will need to correct by the square root of the product of the mapping consistency and internal consistency of the units, in order to properly approximate the true value of $\mathcal{M}$ in the limit of infinite trials.

To make this correction explicit, suppose we have neural responses from two animals (or subjects) $\animalA$ and $\animalB$.
Let $\mathrm{t}_i^p$ be the vector of true responses (either at a given time bin or averaged across a set of time bins) of animal $p \in \mathcal{A} = \{\animalA,\animalB,\dots\}$ on stimulus set $i \in \{\train, \test\}$.
Of course, we only receive noisy observations of $\mathrm{t}_i^p$, so let $\mathrm{s}_{j,i}^p$ be the $j$th set of $n$ trials of $\mathrm{t}_i^p$.
Finally, let $M(x;y)_i$ be the predictions of a mapping $M$ (e.g., PLS, or any type of regression) when trained on input $x$ to match output $y$ and tested on stimulus set $i$.
For example, $M\left(\trueA;\trueB\right)_{\test}$ is the prediction of mapping $M$ on the test set stimuli trained to match the true neural responses of animal $\animalB$ given, as input, the true neural responses of animal $\animalA$ on the train set stimuli.
Similarly, $M\left(\sfAtrain;\sfBtrain\right)_{\test}$ is the prediction of mapping $M$ on the test set stimuli trained to match the trial-average of noisy sample 1 on the train set stimuli of animal $\animalB$ given, as input, the trial-average of noisy sample 1 on the train set stimuli of animal $\animalA$.
Then we have that:
\begin{equation}\label{neuroaittestcorr}
\begin{split}
&\Mtrue := \left\langle \corr\left(M\left(\trueA;\trueB\right)_{\test}, \trueBtest\right)\right\rangle \\
& \sim \Mest := \Biggl\langle
  \dfrac{
    \overbrace{\corr\!\bigl(M(\sfAtrain;\sfBtrain)_{\test},\;\ssBtest\bigr)}^{\text{predictivity}}
  }{
    \sqrt{
      \underbrace{\widetilde{\corr}\!\bigl(M(\sfAtrain;\sfBtrain)_{\test},\;M(\ssAtrain;\ssBtrain)_{\test}\bigr)}_{\text{mapping consistency}}
      \;\times\;
      \underbrace{\widetilde{\corr}\!\bigl(\sfBtest,\;\ssBtest\bigr)}_{\text{internal consistency}}
    }
  }
\Biggr\rangle,
\end{split}
\end{equation}
where the average $\langle\cdot\rangle$ is taken over bootstrapped split-half trials and train-test splits, and $\corr(\cdot,\cdot)$ denotes the Pearson correlation of the two quantities.
$\widetilde{\corr}(\cdot, \cdot)$ denotes the Spearman-Brown corrected value of the original quantity (since it is computed on split-halves of the trials, unlike the numerator, which is evaluated on the full trial set).
The analogous correction for RSA holds, where the RDM/RSM of the responses is instead used for $s$, and $M$ is the identity mapping, $M(x;\cdot)_{\mathsf{test}} \equiv x_{\mathsf{test}}$.
When constructing $\Mest$ for model-brain mappings (rather than brain-brain mappings), we just replace $\animalA$ with the model responses, which are deterministic.

The above correction in \eqref{neuroaittestcorr} is fully implemented in the 
\texttt{brainmodel\_utils} package (\url{https://github.com/neuroagents-lab/brainmodel_utils}), specifically in the \texttt{get\_linregress\_consistency} function. 
This function can be imported as follows:

\begin{lstlisting}
from brainmodel_utils.metrics.consistency import get_linregress_consistency
\end{lstlisting}

The \texttt{r\_xy\_n\_sb} value returned by this function corresponds to the ratio in \eqref{neuroaittestcorr}.
Refer to the \href{https://github.com/neuroagents-lab/brainmodel_utils#readme}{README} and the function docstring for usage details across a range of linearly regressed and non-regressed (e.g. RSA) metrics.

\subsection{Single Subject Pair}
\label{ss:methods-interanimal-pair}
Suppose we have neural responses from two animals (or subjects) $\animalA$ and $\animalB$.
Let $\mathrm{t}_i^p$ be the vector of true responses (either at a given time bin or averaged across a set of time bins) of animal $p \in \mathcal{A} = \{\animalA,\animalB,\dots\}$ on stimulus set $i \in \{\train, \test\}$.
Of course, we only receive noisy observations of $\mathrm{t}_i^p$, so let $\mathrm{s}_{j,i}^p$ be the $j$th set of $n$ trials of $\mathrm{t}_i^p$.
Finally, let $M(x;y)_i$ be the predictions of a mapping $M$ (e.g., PLS) when trained on input $x$ to match output $y$ and tested on stimulus set $i$.
For example, $M\left(\trueA;\trueB\right)_{\test}$ is the prediction of mapping $M$ on the test set stimuli trained to match the true neural responses of animal $\animalB$ given, as input, the true neural responses of animal $\animalA$ on the train set stimuli.
Similarly, $M\left(\sfAtrain;\sfBtrain\right)_{\test}$ is the prediction of mapping $M$ on the test set stimuli trained to match the trial-average of noisy sample 1 on the train set stimuli of animal $\animalB$ given, as input, the trial-average of noisy sample 1 on the train set stimuli of animal $\animalA$.

With these definitions in hand, the inter-animal mapping consistency from animal $\animalA$ to animal $\animalB$ corresponds to the following ``true'' quantity to be estimated by $\Mest$ in the limit of infinite trials:
\begin{equation}\label{interancontrue}
\mathcal{M}_{\text{true}} := \corr\left(M\left(\trueA;\trueB\right)_{\test}, \trueBtest\right),
\end{equation}
where $\corr(\cdot, \cdot)$ is the Pearson correlation across a stimulus set.
In what follows, we will argue that Eq~\eqref{interancontrue} can be approximated with the following ratio of measurable quantities, where we split in half and average the noisy trial observations, indexed by 1 and by 2:
\begin{equation}\label{interancon}
\begin{split}
& \Mtrue := \corr\left(M\left(\trueA;\trueB\right)_{\test}, \trueBtest\right) \\
& \sim \Mest := \dfrac{\corr\left(M\left(\sfAtrain;\sfBtrain\right)_{\test}, \ssBtest\right)}{\sqrt{\corr\left(M\left(\sfAtrain;\sfBtrain\right)_{\test}, M\left(\ssAtrain;\ssBtrain\right)_{\test}\right) \times \corr\left(\sfBtest, \ssBtest\right)}}.
\end{split}
\end{equation}
In words, the inter-animal consistency (i.e., the quantity on the left side of Eq~\eqref{interancon}) corresponds to the predictivity of the mapping on the test set stimuli from animal $\animalA$ to animal $\animalB$ on two different (averaged) halves of noisy trials (i.e., the numerator on the right side of Eq~\eqref{interancon}), corrected by the square root of the mapping reliability on animal $\animalA$'s responses to the test set stimuli on two different halves of noisy trials multiplied by the internal consistency of animal $\animalB$.

We justify the approximation in Eq~\eqref{interancon} by gradually replacing the true quantities ($\mathrm{t}$) by their measurable estimates ($\mathrm{s}$), starting from the original quantity in Eq~\eqref{interancontrue}.
First, we make the approximation that:
\begin{equation}\label{step1}
\corr\left(M\left(\trueA;\trueB\right)_{\test}, \ssBtest\right) \sim \corr\left(M\left(\trueA;\trueB\right)_{\test}, \trueBtest\right) \times \corr\left(\trueBtest, \ssBtest\right),
\end{equation}
by the transitivity of very positive correlations. 
Namely, in scenarios where correlations are very close to 1, a form of transitivity holds, meaning if variable $A$ is highly correlated with variable $B$, and variable $B$ with variable $C$, then variable $A$ is also highly correlated with variable $C$. 
This is the desired situation, as low or negative correlations indicate neurons that are not self-consistent.
Moreover, calculating certain metrics in these cases can result in undefined values due to operations like taking the square root of a negative number. 
Assuming high correlations is reasonable, especially when the number of stimuli is large.
Next, by transitivity and normality assumptions in the structure of the noisy estimates and since the number of trials ($n$) between the two sets is the same, we have that:
\begin{align}\label{step2}
\corr\left(\sfBtest, \ssBtest\right) &\sim \corr\left(\sfBtest, \trueBtest\right) \times \corr\left(\trueBtest, \ssBtest\right) \nonumber \\
&\sim \corr\left(\trueBtest, \ssBtest\right)^2.
\end{align}
In words, Eq~\eqref{step2} states that the correlation between the average of two sets of noisy observations of $n$ trials each is approximately the square of the correlation between the true value and average of one set of $n$ noisy trials.
Therefore, combining Eq~\eqref{step1} and Eq~\eqref{step2}, it follows that:
\begin{equation}\label{lemma1}
\corr\left(M\left(\trueA;\trueB\right)_{\test}, \trueBtest\right) \sim \dfrac{\corr\left(M\left(\trueA;\trueB\right)_{\test}, \ssBtest\right)}{\sqrt{\corr\left(\sfBtest, \ssBtest\right)}}.
\end{equation}

From the right side of Eq~\eqref{lemma1}, we can see that we have removed $\trueBtest$, but we still need to remove the $M\left(\trueA;\trueB\right)_{\test}$ term, as this term still contains unmeasurable (i.e., true) quantities.
We apply the same two steps, described above, by analogy, though these approximations may not always be true (they are, however, true for Gaussian noise):
\begin{equation*}
\begin{split}
\corr\left(M\left(\sfAtrain;\sfBtrain\right)_{\test}, \ssBtest\right) & \sim \corr\left(\ssBtest, M\left(\trueA;\trueB\right)_{\test}\right) \\
& \times \corr\left(M\left(\trueA;\trueB\right)_{\test}, M\left(\sfAtrain;\sfBtrain\right)_{\test}\right)
\end{split}
\end{equation*}
\begin{equation*}
\begin{split}
& \corr\left(M\left(\sfAtrain;\sfBtrain\right)_{\test}, M\left(\ssAtrain;\ssBtrain\right)_{\test}\right) \\
& \sim \corr\left(M\left(\sfAtrain;\sfBtrain\right)_{\test}, M\left(\trueA;\trueB\right)_{\test}\right)^2,
\end{split}
\end{equation*}
which taken together implies the following:
\begin{equation}\label{lemma2}
\corr\left(M\left(\trueA;\trueB\right)_{\test}, \ssBtest\right) \sim \dfrac{\corr\left(M\left(\sfAtrain;\sfBtrain\right)_{\test}, \ssBtest\right)}{\sqrt{\corr\left(M\left(\sfAtrain;\sfBtrain\right)_{\test}, M\left(\ssAtrain;\ssBtrain\right)_{\test}\right)}}.
\end{equation}
Eq~\eqref{lemma1} and Eq~\eqref{lemma2} together imply the final estimated quantity given in Eq~\eqref{interancon}.

\subsection{Multiple Subject Pairs}
\label{ss:methods-interanimal-multiple}
For multiple animals, we consider the average of the true quantity for each target in $\animalB$ in Eq~\eqref{interancontrue} across source animals $\animalA$ in the ordered pair $(\animalA,\animalB)$ of animals $\animalA$ and $\animalB$:
\begin{equation*}\label{multipleinterancon}
\begin{split}
&\Mtrue := \left\langle \corr\left(M\left(\trueA;\trueB\right)_{\test}, \trueBtest\right)\right\rangle_{\animalA \in \mathcal{A}: (\animalA,\animalB)\in \mathcal{A}\times\mathcal{A}} \\
& \sim \Mest := \left\langle\dfrac{\corr\left(M\left(\sfAtrain;\sfBtrain\right)_{\test}, \ssBtest\right)}{\sqrt{\widetilde{\corr}\left(M\left(\sfAtrain;\sfBtrain\right)_{\test}, M\left(\ssAtrain;\ssBtrain\right)_{\test}\right) \times \widetilde{\corr}\left(\sfBtest, \ssBtest\right)}}\right\rangle_{\animalA \in \mathcal{A}: (\animalA,\animalB)\in \mathcal{A}\times\mathcal{A}}.
\end{split}
\end{equation*}
We also bootstrap across trials, and have multiple train/test splits, in which case the average on the right hand side of the equation includes averages across these as well.

Note that each neuron in our analysis will have this single average value associated with it when \emph{it} was a target animal ($\animalB$), averaged over source animals/subsampled source neurons, bootstrapped trials, and train/test splits.
This yields a vector of these average values, which we can take median and standard error of the mean (s.e.m.) over, as we do with standard explained variance metrics.

\subsection{RSA}
\label{ss:methods-interanimal-rsa}
We can extend the above derivations to other commonly used metrics for comparing representations that involve correlation.
Since $\rsa(x,y) \coloneqq \corr(\rdm(x), \rdm(y))$, then the corresponding quantity in Eq~\eqref{interancon} analogously (by transitivity of maximally positive correlations) becomes:
\begin{equation}\label{rsainterancon}
\begin{split}
&\Mtrue := \left\langle\rsa\left(M\left(\trueA;\trueB\right)_{\test}, \trueBtest\right)\right\rangle_{\animalA \in \mathcal{A}: (\animalA,\animalB)\in \mathcal{A}\times\mathcal{A}} \\
& \sim \Mest := \left\langle\dfrac{\rsa\left(M\left(\sfAtrain;\sfBtrain\right)_{\test}, \ssBtest\right)}{\sqrt{\widetilde{\rsa}\left(M\left(\sfAtrain;\sfBtrain\right)_{\test}, M\left(\ssAtrain;\ssBtrain\right)_{\test}\right) \times \widetilde{\rsa}\left(\sfBtest, \ssBtest\right)}}\right\rangle_{\animalA \in \mathcal{A}: (\animalA,\animalB)\in \mathcal{A}\times\mathcal{A}}.
\end{split}
\end{equation}

Note that in this case, each \emph{animal} (rather than neuron) in our analysis will have this single average value associated with it when \emph{it} was a target animal ($\animalB$) (since RSA is computed over images and neurons), where the average is over source animals/subsampled source neurons, bootstrapped trials, and train/test splits.
This yields a vector of these average values, which we can take median and s.e.m. over, across animals $\animalB \in \mathcal{A}$.

For RSA, we can use the identity mapping (since RSA is computed over neurons as well, the number of neurons between source and target animal can be different to compare them with the identity mapping). 
As parameters are not fit, we can choose $\train = \test$, so that Eq~\eqref{rsainterancon} becomes:
\begin{equation}\label{rsainteranconid}
\Mtrue := \left\langle\rsa\left(\trueAid,\trueBid\right)\right\rangle_{\animalA \in \mathcal{A}: (\animalA,\animalB)\in \mathcal{A}\times\mathcal{A}} \sim \Mest := \left\langle\dfrac{\rsa\left(\sfAid, \ssBid\right)}{\sqrt{\widetilde{\rsa}\left(\sfAid, \ssAid\right) \times \widetilde{\rsa}\left(\sfBid, \ssBid\right)}}\right\rangle_{\animalA \in \mathcal{A}: (\animalA,\animalB)\in \mathcal{A}\times\mathcal{A}}.
\end{equation}

\subsection{Pooled Source Animal}
\label{ss:methods-interanimal-holdouts}
Often times, we may not have enough neurons per animal to ensure that the estimated inter-animal consistency in our data closely matches the ``true'' inter-animal consistency.
In order to address this issue, we holdout one animal at a time and compare it to the pseudo-population aggregated across units from the remaining animals, as opposed to computing the consistencies in a pairwise fashion.
Thus, $\animalB$ is still the target heldout animal as in the pairwise case, but now the average over $\animalA$ is over a sole ``pooled'' source animal constructed from the pseudo-population of the remaining animals.

Pooling data across subjects to create larger pseudopopulations is a common practice, and helps researchers better isolate core representational principles that are conserved across individuals when data collection modalities limit the number of collected neurons per session.

\subsection{Spearman-Brown Correction}
\label{ss:methods-interanimal-spearman-brown}
The Spearman-Brown correction can be applied to each of the terms in the denominator individually, as they are each correlations of observations from half the trials of the \emph{same} underlying process to itself (unlike the numerator). Namely,
\begin{equation*}
\widetilde{\corr}\left(X,Y\right) \coloneqq \frac{2\corr\left(X,Y\right)}{1 + \corr\left(X,Y\right)}.
\end{equation*}
Analogously, since $\rsa(X,Y) \coloneqq \corr(\rdm(x), \rdm(y))$, then we define
\begin{align*}
\widetilde{\rsa}\left(X,Y\right) &\coloneqq \widetilde{\corr}(\rdm(x), \rdm(y)) \\
    &= \frac{2\rsa\left(X,Y\right)}{1 + \rsa\left(X,Y\right)}.
\end{align*}

\newpage
\section*{NeurIPS Paper Checklist}

The checklist is designed to encourage best practices for responsible machine learning research, addressing issues of reproducibility, transparency, research ethics, and societal impact. Do not remove the checklist: {\bf The papers not including the checklist will be desk rejected.} The checklist should follow the references and follow the (optional) supplemental material.  The checklist does NOT count towards the page
limit. 

Please read the checklist guidelines carefully for information on how to answer these questions. For each question in the checklist:
\begin{itemize}
    \item You should answer \answerYes{}, \answerNo{}, or \answerNA{}.
    \item \answerNA{} means either that the question is Not Applicable for that particular paper or the relevant information is Not Available.
    \item Please provide a short (1–2 sentence) justification right after your answer (even for NA). 
\end{itemize}

{\bf The checklist answers are an integral part of your paper submission.} They are visible to the reviewers, area chairs, senior area chairs, and ethics reviewers. You will be asked to also include it (after eventual revisions) with the final version of your paper, and its final version will be published with the paper.

The reviewers of your paper will be asked to use the checklist as one of the factors in their evaluation. While "\answerYes{}" is generally preferable to "\answerNo{}", it is perfectly acceptable to answer "\answerNo{}" provided a proper justification is given (e.g., "error bars are not reported because it would be too computationally expensive" or "we were unable to find the license for the dataset we used"). In general, answering "\answerNo{}" or "\answerNA{}" is not grounds for rejection. While the questions are phrased in a binary way, we acknowledge that the true answer is often more nuanced, so please just use your best judgment and write a justification to elaborate. All supporting evidence can appear either in the main paper or the supplemental material, provided in appendix. If you answer \answerYes{} to a question, in the justification please point to the section(s) where related material for the question can be found.

IMPORTANT, please:
\begin{itemize}
    \item {\bf Delete this instruction block, but keep the section heading ``NeurIPS Paper Checklist"},
    \item  {\bf Keep the checklist subsection headings, questions/answers and guidelines below.}
    \item {\bf Do not modify the questions and only use the provided macros for your answers}.
\end{itemize}


\begin{enumerate}

\item {\bf Claims}
    \item[] Question: Do the main claims made in the abstract and introduction accurately reflect the paper's contributions and scope?
    \item[] Answer: \answerYes{} 
    \item[] Justification: We confirm that the claims made in the abstract and introduction do not exaggerate or fabricate the scope of our contributions and results. 

\item {\bf Limitations}
    \item[] Question: Does the paper discuss the limitations of the work performed by the authors?
    \item[] Answer: \answerYes{} 
    \item[] Justification: We confirm that we include a paragraph on limitations in the conclusions section. 

\item {\bf Theory assumptions and proofs}
    \item[] Question: For each theoretical result, does the paper provide the full set of assumptions and a complete (and correct) proof?
    \item[] Answer: \answerNA{} 
    \item[] Justification: We do not provide any theoretical results in our paper.

    \item {\bf Experimental result reproducibility}
    \item[] Question: Does the paper fully disclose all the information needed to reproduce the main experimental results of the paper to the extent that it affects the main claims and/or conclusions of the paper (regardless of whether the code and data are provided or not)?
    \item[] Answer: \answerYes{}
    \item[] Justification: We fully detail the architectures, environment, and algorithms used, as well as the data used for model-brain alignment comparisons. 

\item {\bf Open access to data and code}
    \item[] Question: Does the paper provide open access to the data and code, with sufficient instructions to faithfully reproduce the main experimental results, as described in supplemental material?
    \item[] Answer: \answerYes{} 
    \item[] Justification: We will provide full access to our code, as well as links to access the larval zebrafish data. 

\item {\bf Experimental setting/details}
    \item[] Question: Does the paper specify all the training and test details (e.g., data splits, hyperparameters, how they were chosen, type of optimizer, etc.) necessary to understand the results?
    \item[] Answer:  \answerYes{} 
    \item[] Justification: We provide a complete characterization of all the methods in our paper, with more details in the supplemental materials. 

\item {\bf Experiment statistical significance}
    \item[] Question: Does the paper report error bars suitably and correctly defined or other appropriate information about the statistical significance of the experiments?
    \item[] Answer: \answerYes{} 
    \item[] Justification: We report error bars on all our plots when applicable.

\item {\bf Experiments compute resources}
    \item[] Question: For each experiment, does the paper provide sufficient information on the computer resources (type of compute workers, memory, time of execution) needed to reproduce the experiments?
    \item[] Answer: \answerYes{} 
    \item[] Justification: We include details on the computational resources we used. 
    
\item {\bf Code of ethics}
    \item[] Question: Does the research conducted in the paper conform, in every respect, with the NeurIPS Code of Ethics \url{https://neurips.cc/public/EthicsGuidelines}?
    \item[] Answer: \answerYes{} 
    \item[] Justification: We strickly adhere to the NeurIPS code of Ethics. 

\item {\bf Broader impacts}
    \item[] Question: Does the paper discuss both potential positive societal impacts and negative societal impacts of the work performed?
    \item[] Answer: \answerNA{} 
    \item[] Justification: Our paper addressed fundamental scientific questions, and is not directly aimed towards societal impact. 
    
\item {\bf Safeguards}
    \item[] Question: Does the paper describe safeguards that have been put in place for responsible release of data or models that have a high risk for misuse (e.g., pretrained language models, image generators, or scraped datasets)?
    \item[] Answer: \answerNA{} 
    \item[] Justification: Our paper does not have  high risk for misuse. 

\item {\bf Licenses for existing assets}
    \item[] Question: Are the creators or original owners of assets (e.g., code, data, models), used in the paper, properly credited and are the license and terms of use explicitly mentioned and properly respected?
    \item[] Answer: \answerYes{} 
    \item[] Justification: We make sure to properly credit everybody. 

\item {\bf New assets}
    \item[] Question: Are new assets introduced in the paper well documented and is the documentation provided alongside the assets?
    \item[] Answer: \answerNA{} 
    \item[] Justification: We do not provide any new assests.

\item {\bf Crowdsourcing and research with human subjects}
    \item[] Question: For crowdsourcing experiments and research with human subjects, does the paper include the full text of instructions given to participants and screenshots, if applicable, as well as details about compensation (if any)? 
    \item[] Answer: \answerNA{} 
    \item[] Justification: We do not conduct any crowdsourcing experiments. 

\item {\bf Institutional review board (IRB) approvals or equivalent for research with human subjects}
    \item[] Question: Does the paper describe potential risks incurred by study participants, whether such risks were disclosed to the subjects, and whether Institutional Review Board (IRB) approvals (or an equivalent approval/review based on the requirements of your country or institution) were obtained?
    \item[] Answer: \answerNA{} 
    \item[] Justification: Our paper does not use participants. 

\item {\bf Declaration of LLM usage}
    \item[] Question: Does the paper describe the usage of LLMs if it is an important, original, or non-standard component of the core methods in this research? Note that if the LLM is used only for writing, editing, or formatting purposes and does not impact the core methodology, scientific rigorousness, or originality of the research, declaration is not required.
    \item[] Answer: \answerNA{} 
    \item[] Justification: We do not use LLMs for the mentioned purposes. 

\end{enumerate}
\end{document}